\documentclass[prd,amsfonts,onecolumn,superscriptaddress,aps,nofootinbib,11pt]{revtex4-1}

\usepackage[top=3cm,bottom=3cm,left=2cm,right=2cm,marginparwidth=1.75cm]{geometry}

\usepackage{amssymb,amsmath}
\usepackage{subfig}
\usepackage{graphicx}
\usepackage{xcolor}
\usepackage{hyperref}
\usepackage{listings}
\usepackage{soul}
\hypersetup{
    unicode=false,          
    pdftoolbar=true,        
    pdfmenubar=true,        
    pdffitwindow=false,     
    pdfauthor={William},     
    colorlinks=true,       
    linkcolor=blue,          
    citecolor=red,        
    urlcolor=blue
}

\graphicspath{{figures/}}

\providecommand{\be}{ \begin{equation} }
\providecommand{\ee}{\end{equation}}
\providecommand{\bea}{\begin{eqnarray}}
\providecommand{\eea}{\end{eqnarray}}
\providecommand{\nn}{\nonumber}

\def\331{SU(3)_C\otimes SU(3)_L\otimes U(1)_X}
\def\3311{$\text{SU}(3)_C \otimes\text{SU}(3)_L\otimes\text{U}(1)_X \otimes \text{U}(1)_N$}

\definecolor{darkpastelpurple}{rgb}{0.59, 0.44, 0.84}
\definecolor{frenchlilac}{rgb}{0.53, 0.38, 0.56}
\definecolor{violet}{rgb}{0.56, 0.0, 1.0}

\usepackage{hyperref}
\hypersetup{
     colorlinks = true,
     linkcolor = violet,
     urlcolor  = blue,
     citecolor = magenta,
     anchorcolor = cyan,
}

\usepackage{amsmath}

\begin{document}

\lstset{frame=tb,
  	language=Matlab,
  	aboveskip=3mm,
  	belowskip=3mm,
 	showstringspaces=false,
	columns=flexible,
  	basicstyle={\small\ttfamily},
  	numbers=none,
  	numberstyle=\tiny\color{gray},
 	keywordstyle=\color{blue},
	commentstyle=\color{green},
  	stringstyle=\color{mauve},
  	breaklines=true,
  	breakatwhitespace=true
  	tabsize=3
}

\title{Scale-invariant 3-3-1-1 model with $B-L$ symmetry}
\author{Alex G. Dias}
\email{alex.dias@ufabc.edu.br}
\affiliation{Centro de Ci\^encias Naturais e Humanas, Universidade Federal do ABC,\\
09210-580, Santo Andr\'e-SP, Brasil}
\author{Julio Leite}
\email{julio.leite@ific.uv.es}
\affiliation{AHEP Group, Institut de F\'{i}sica Corpuscular --
  C.S.I.C./Universitat de Val\`{e}ncia, Parc Cient\'ific de Paterna.\\
 C/ Catedr\'atico Jos\'e Beltr\'an, 2 E-46980 Paterna (Valencia) - Spain}
\author{B. L. S\'anchez-Vega}
\email{bruce@fisica.ufmg.br}
\affiliation{Departamento de F\'isica, UFMG, Belo Horizonte, MG 31270-901, Brasil.}
\date{\today}

\begin{abstract}

Motivated by a possible interplay between the mechanism of  dynamical symmetry breaking and the  seesaw mechanism for generating fermion masses, we present a scale-invariant model that extends the gauge symmetry of the Standard Model electroweak sector to SU(3)$_L\otimes$U(1)$_X\otimes$U(1)$_N$, with a built-in $B-L$ symmetry. The model is based on the symmetry structure of the known 3-3-1 models and, thus, it relates the number of the three observed fermion generations with the cancellation of gauge anomalies. Symmetry breaking is triggered via the Coleman-Weinberg mechanism taking into account a minimal set of scalar field multiplets. We establish the stability conditions for the tree-level scalar potential imposing the copositivity criteria and use the method of Gildener-Weinberg for computing the one-loop effective potential when one has multiple scalar fields. With the addition of vectorial fermions, getting their mass mainly through the vacuum expectation value of scalar singlets at $10^3$ TeV, the $B-L$ symmetry leads to textures for the fermion mass matrices, allowing seesaw mechanisms for neutrinos and quarks to take place. In particular, these mechanisms could partly explain the mass hierarchies of the quarks. Once the breakdown of the SU(3)$_L$ symmetry is supposed to occur  around 10 TeV, the model also predicts new particles with TeV-scale masses, such as a neutral scalar, $H_{1}$, a charged scalar, $H^\pm$, and the gauge bosons $Z^{\prime}$, $W^{\prime\pm}$ and $Y^0$, that could be searched with the high-luminosity LHC.

\end{abstract}

\maketitle

\section{Introduction}

It is plausible that the conventional spontaneous symmetry breaking  mechanism in the Standard Model (SM) has a dynamical origin. The seminal proposal of the Coleman-Weinberg (CW) mechanism is a way to trigger spontaneous symmetry breakdown dynamically through radiative corrections to the scalar potential~\cite{Coleman:1973jx,Weinberg:1973am}. It assumes that the Lagrangian has a classical scale invariance -- {\it i.e.}, the absence of any initial energy scales in the theory -- in such a way that an effective potential with a nontrivial minimum can be generated, leading to a vacuum expectation value (vev) for scalar fields and, consequently, symmetry breaking. The emergence of such an energy scale happens in the replacement of one of the coupling constants, through the potential minimisation, known as dimensional transmutation. In principle, this leads to a prediction for the mass of the scalon -- the pseudo-Goldstone boson of the scale symmetry -- in terms of the masses of the other particles.    

The implementation of the CW mechanism at the one-loop level in the SM,  {\it i.e.}, in replacement of its conventional spontaneous symmetry breaking mechanism, results in an unstable potential due to the dominant-negative contribution from the top quark. Still, there are results indicating that the potential is stable when higher-loop corrections are included, but it has yet to be shown that the Higgs boson mass converges to the measured value~\cite{Elias:2003zm,Elias:2003xp,Chishtie:2010ni,Steele:2012av}. Anyway, the idea of dynamical symmetry breaking via the CW mechanism has motivated the construction of a variety of models. It may be the way of solving the hierarchy problem once, with this symmetry, the quantum corrections are just logarithmic in the fields for the effective potential~\cite{Bardeen:1995kv}, whose stability could be taken for granted until the Planck scale in certain SM extensions~\cite{Hempfling:1996ht,Meissner:2006zh}. The CW mechanism has been considered in extensions of the SM involving, for example, scalar singlets and right-handed neutrinos for seesaw mechanisms~\cite{Dias:2006th,Meissner:2006zh,Iso:2009ss,Iso:2009nw,Karam:2015jta, Das:2015nwk, Das:2016zue}, the break of global symmetries related to lepton number, axions and axion-like particles~\cite{Dias:2006th,Dias:2005jk,Meissner:2008gj,Iso:2012jn,Bertolini:2015boa}, and dark matter~\cite{Foot:2010av,Gabrielli:2013hma,Steele:2013fka,Altmannshofer:2014vra, Karam:2015jta,Plascencia:2015xwa,Latosinski:2015pba,Wang:2015cda,Ahriche:2015loa, Ghorbani:2015xvz,Helmboldt:2016mpi,Khoze:2016zfi,Karam:2016rsz,Oda:2017kwl,Hambye:2018qjv,YaserAyazi:2019caf,Kannike:2022pva}. 

In this work, we present a scale-invariant model which extends the SM symmetry of the electroweak sector to SU(3)$_L\otimes$U(1)$_X\otimes$U(1)$_{N}$~\cite{Dong:2013wca,Dong:2015yra,Alves:2016fqe,Leite:2019grf,Leite:2020bnb}. The embedding of the SM fermionic fields in the symmetry multiplets is such that the gauge anomalies are cancelled for an integer multiple of three fermion generations, as originally proposed in the models with SU(3)$_L\otimes$U(1)$_X$ symmetry, known as 3-3-1 models~\cite{Singer:1980sw,Pisano:1991ee,Foot:1992rh,Frampton:1992wt,Montero:1992jk,Pleitez:1992xh,Foot:1994ym,Pleitez:1994pu,Ozer:1995xi,Hoang:1995vq}. This could be a hint to the fermion generation replication puzzle (for a recent discussion on 3-3-1 model aspects see~\cite{Pleitez:2021abk}). The model features an interplay of dynamical symmetry breaking and seesaw-type mechanisms for the masses of quarks and neutrinos. The construction takes into account a minimal set of scalar fields, composed of two SU(3)$_L$ triplets, a complex singlet, and a real singlet, as well as vectorial multiplets of quarks and neutral leptons. The extra fermions get their masses from interactions with the scalar singlets, setting up -- along with a built-in $B-L$ symmetry -- seesaw textures for the fermion mass matrices. We show in detail how the dynamically generated  vevs of the scalar fields can give rise to mass hierarchies for the quarks, which could partly explain the flavour puzzle -- the unexplained values of fermion masses and mixing angles in the SM. 

In dealing with multiple scalar fields, we apply the method of Gildener-Weinberg to compute the effective potential~\cite{Gildener:1976ih}. It is based on the assumption that there is an energy scale where the coupling constants allow for a flat direction in the tree-level potential, where the one-loop effective potential is computed (for a review of dynamical symmetry breaking in scale-invariant models, see \cite{Helmboldt:2016mpi,AlexanderNunneley:2010nw}). Before performing this analysis, we study the tree-level potential stability with the imposition of the copositivity criteria on the matrix of couplings~\cite{Kannike:2012pe,Kannike:2016fmd,Kannike:2019upf}. 

In the end, the scalar particle spectrum is composed of four neutral CP-even scalars -- one of them being the scalon -- and a charged scalar. Among the neutral scalar bosons, only one has its mass proportional to the electroweak scale and, thus, is naturally identified with the discovered Higgs boson. All the remaining scalar bosons have masses directly proportional to the new energy scales, which could be around or above a few TeVs. This scalar particle spectrum is simpler than those in the typical 3-3-1 models, which assume at least three scalar triplets to break the symmetries with the conventional mechanism. It is also in agreement with the current searches for fundamental scalars once no evidence for new scalar particles beyond the Higgs boson has been found yet, indicating that, if they exist, these particles should be related to new high-energy scales.

The present study follows the first work on dynamical symmetry breaking in the context of 3-3-1 models~\cite{Dias:2020ryz}, recently proposed by the authors. Building upon what was learnt then, we present an improved construction based on another 3-3-1 version, often referred to as the 3-3-1 model with right-handed neutrinos. Amongst the main differences between the two studies, we point out the presence, in the current proposal, of a gauged $B-L$ symmetry, arising in an elegant fashion, which helps to impose a seesaw texture for the fermion mass matrices and to stabilise the effective potential.

The work is organized as follows. In section~\ref{scainv331} we present pedagogically the construction of our model starting from a scale-invariant 3-3-1 model and its limitations, as well as the emergence of a $B-L$ symmetry and its role. In section~\ref{scsect} we work out the scale-invariant potential, the stability conditions, and the flat directions for applying the Gildener-Weinberg method. We also discuss the scalar particle spectrum, identifying the scalon field and the Higgs boson. The gauge bosons are discussed in section~\ref{sec:gauge}. In section~\ref{fermionmassessection} we turn to the fermion masses, presenting our seesaw mechanism for generating hierarchical masses to the quarks and the neutral leptons. The one-loop effective potential, along with the role of the new particles in its stabilisation, is discussed in section~\ref{sec:disc}. Our conclusions are presented in section~\ref{sec:conc}.

\section{The scale-invariant 3-3-1-1 model with a built-in $B-L$ symmetry} 
\label{scainv331}

Our model is presented gradually in this section. First, we discuss the minimal scale-invariant 3-3-1 model, upon which our proposal is built, and highlight some of its limitations. Second, we show that, in such a context, an anomaly-free $B-L$ symmetry can emerge as a residual symmetry. Finally, we propose an extension of the minimal framework that tackles the latter's main drawbacks via dynamical symmetry breaking.

\subsection{The minimal model and its shortcomings}\label{minimal}

Our starting point is a scale-invariant model  based on the $\textrm{SU}\left(3\right)_{C}\otimes\textrm{SU}\left(3\right)_{L}\otimes\textrm{U}\left(1\right)_{X}$ gauge symmetry -- where $C$ stands for colour and $L$ for left chirality as in the SM, whereas $X$ stands for a new charge which is chosen to correctly reproduce the SM hypercharge -- on which the well-known 3-3-1 models are based~\cite{Singer:1980sw,Pisano:1991ee,Foot:1992rh,Frampton:1992wt,Montero:1992jk,Pleitez:1992xh,Foot:1994ym,Pleitez:1994pu,Ozer:1995xi,Hoang:1995vq}. It is instructive to review some aspects of the model building which have motivated our construction. 

The electric charge operator defining the charges of the field components is 
\begin{equation}\label{Q}
    Q=T_{3}-\frac{1}{\sqrt{3}}T_{8}+X\,\mathbf{1}_{3\times3},
\end{equation}
where $T_{3}$ and $T_{8}$ are the diagonal SU$\left(3\right)_{L}$ generators. The new charge $X$ is closely related to the SM extended hypercharge operator $Y= -\frac{2}{\sqrt{3}}T_{8}+2X\,\mathbf{1}_{3\times3}$. Note that for a field in the fundamental representation of $\textrm{SU}\left(3\right)_{L}$, $\mathbf{3}$, the electric charges of its components are $(X+1/3,\,X-2/3,\,X+1/3)$, respectively, whereas in the anti-fundamental representation, $\bar{\mathbf{3}}$, the electric charges of its components are $(X-1/3,\,X+2/3,\,X-1/3)$. 

A peculiar feature of 3-3-1 models is that the cancellation of gauge anomalies requires a set of fermionic multiplets organised in an integer multiple of three families. This offers a possible explanation for the fact that just three families of fermions have been observed until now. We then consider the minimal set of fermionic multiplets containing the SM fields similar to the 3-3-1 models in Refs.~\cite{Singer:1980sw,Montero:1992jk,Foot:1994ym}. The SM lepton fields turn out to be part of the following multiplets, along with three right-handed neutrinos, $N_{iR}$,  
\begin{eqnarray}
& & F_{iL}=\left(\nu_{i},\,e_{i},\,N_{i}^{c}\right)_{L}^{\textrm{T}}\sim\left(\mathbf{1},\,\mathbf{3},\,-1/3\right), \nonumber\\
& & e_{iR}\sim\left(\mathbf{1},\,\mathbf{1},\,-1\right),
\label{eq:1}
\end{eqnarray} 
where $i=1,2,3$, $c$ stands for the charge conjugation operator, and the numbers in parentheses denote the transformation properties under the local symmetry groups, $\textrm{SU}(3)_{C}$, $\textrm{SU}(3)_{L}$ and $\textrm{U}(1)_{X}$, respectively. The SM quarks, along with a new quark $U$ and two quarks $D_a$, are disposed in left- and right-handed multiplets according to 
\begin{eqnarray}
&& Q_{1L} =\left(u_{1},\,d_{1},\,U\right)_{L}^{\textrm{T}}\sim\left(\mathbf{3},\,\mathbf{3},\,1/3\right), \label{eq:2}\nonumber\\
&& Q_{aL} =\left(d_{a},\,u_{a},\,D_{a}\right)_{L}^{\textrm{T}}\sim\left(\mathbf{3},\,\bar{\mathbf{3}},\,0\right), \\ 
&& u_{sR} \sim\left(\mathbf{3},\,\mathbf{1},\,2/3\right),\quad d_{tR}\sim\left(\mathbf{3},\,\mathbf{1},\,-1/3\right),
\label{eq:3}
\end{eqnarray}
where $a=2,\,3$, $s=1,\dots,4$ and $t=1,\dots,5$. For convenience we define $u_{4R}\equiv U_R$ and $d_{(4,5)R}\equiv D_{(2,3)R}$. 

With just two scalar triplets, 
\begin{eqnarray}
& & \rho=\left(\begin{array}{ccc} \rho^{+}_{1},\, \rho^{0}_{2},\, \rho^{+}_{3} \end{array}\right)^\textrm{T} \sim \left(\mathbf{1},\,\mathbf{3},\,2/3\right), \nonumber\\
& & \chi=\left(\begin{array}{c} \chi^{0}_{1},\, \chi^{-}_{2},\, \chi^{0}_{3} \end{array}\right)^\textrm{T}\sim \left(\mathbf{1},\:\mathbf{3},\,-1/3\right),
\end{eqnarray}
acquiring vacuum expectation value (vev) $\langle \rho\rangle=(\begin{array}{ccc} 0,\, v_\rho/\sqrt2,\, 0 \end{array})^\textrm{T}$ and $\langle \chi\rangle=(\begin{array}{c} 0,\, 0,\, v_\chi/\sqrt2 \end{array})^\textrm{T}$ one can break the $\textrm{SU}\left(3\right)_{L}\otimes\textrm{U}\left(1\right)_{X}$ symmetry down to the electromagnetic factor $\textrm{U}\left(1\right)_{Q}$. The configuration in which both neutral components of the triplet $\chi$ get vevs is equivalent to the one displayed above, once one can rotate  $\left\langle  \chi \right\rangle =\left(\left\langle \chi_1^0 \right\rangle,\, 0,\, \left\langle \chi_3^0 \right\rangle\right)$ through a SU$(3)_L$ transformation given by $\exp(i\, \omega_5 T_5)$, with $\omega_5=-2\,\arctan(\left\langle \chi_1^0 \right\rangle/\left\langle \chi_3^0 \right\rangle)$, setting the vev of $\chi^{0}_{1}$ to zero. Also, given that $\langle \chi\rangle$ breaks $\textrm{SU}\left(3\right)_{L}\otimes\textrm{U}\left(1\right)_{X}$ to the SM    $\textrm{SU}\left(2\right)_{L}\otimes\textrm{U}\left(1\right)_{Y}$ symmetry, it is natural to assume $v_\chi>v_\rho$. 

The Yukawa Lagrangian for the leptons involves the triplet $\rho$ and is given by 
\begin{eqnarray}
-\mathcal{L}_{l}^Y=Y^{e}_{ij}\overline{F_{iL}}e_{jR}\rho+Y^{\nu}_{ij}\overline{F_{iL}}\left(F_{jL}\right)^{c}\rho^{*}+ \textrm{H.c.}, 
\label{yuklel}
\end{eqnarray}
where $Y^{e}$ is a $3\times 3$ complex matrix, and $Y^{\nu}$ is an anti-symmetric matrix.  Furthermore, the term $Y^{\nu}_{ij}\overline{F_{iL}}\left(F_{jL}\right)^{c}\rho^{*}$ which contains three SU$(3)_L$ triplets is implicitly contracted with the totally anti-symmetric tensor $\epsilon_{\alpha\beta\gamma}$ ($\alpha,\beta,\gamma$ are SU(3)$_L$ indices), a convention we will follow from here on. By replacing $\rho$ with its vev, one can easily verify that the first term leads to three massive charged leptons. On the other hand, the anti-symmetric nature of the second term, despite resembling a Majorana mass term, leads to mixing between the first and the third neutral components of the lepton triplets, resulting in only two massive -- and degenerate -- Dirac neutrinos, while the other neutral leptons remain massless. Therefore, such a neutrino spectrum needs to be modified since the current findings of oscillation experiments require at least two neutrinos to be massive and necessarily non-degenerate. 

For the quark fields, the Yukawa Lagrangian is 
\begin{eqnarray}
-\mathcal{L}_{\textrm{q}}^{Y}&=& Y^{d}_{1i}\overline{Q_{1L}}d_{iR}\rho+ Y^{D}_{1a}\overline{Q_{1L}}D_{aR}\rho+Y^{u}_{ai}\overline{Q_{aL}}u_{iR}\rho^{*}+Y^{U}_{a}\overline{Q_{aL}}U_{R}\rho^{*}\nonumber\\
&+&Y_{1i}^{u}\overline{Q_{1L}}u_{iR}\chi +Y^{U}_{1}\overline{Q_{1L}}U_{R}\chi+
Y_{ai}^{d}\overline{Q_{a L}}d_{iR}\chi^{\ast}+Y^{D}_{aa'}\overline{Q_{a L}} D_{a' R}\chi^{\ast}+\textrm{H.c.}.
\label{yukq}
\end{eqnarray}
These operators are, however, not enough to generate masses to all quarks. Notably, one up- and two down-type quarks remain massless after spontaneous symmetry breaking takes place.

The renormalisable scale-invariant scalar potential -- quadratic and cubic terms are forbidden -- is given by 
\begin{equation}
V_0 =\lambda_{1}\left(\chi^{\dagger}\chi\right)^{2}+\lambda_{2}\left(\rho^{\dagger}\rho\right)^{2}+\lambda_{3}\left(\chi^{\dagger}\chi\right)\left(\rho^{\dagger}\rho\right)+\lambda_{4}\left(\chi^{\dagger}\rho\right)\left(\rho^{\dagger}\chi\right)\text{.}
\label{v0}
\end{equation}
Assuming that the radiative corrections to this potential would give rise to an effective potential, through the Coleman-Weinberg mechanism, leading to the vevs for the neutral scalar fields and breaking consistently the gauge symmetries, as we have already pointed out, it would not be possible to generate a consistent mass spectrum for the fermions.

The existence of massless particles in the model above is related to an accidental chiral and anomalous symmetry -- it has a colour anomaly -- which we denote as U$(1)_\textrm{PQ}$, a sort of Peccei-Quinn symmetry. The key point is that the scalar sector is not enough to totally break down this chiral symmetry, {\it{i.e.}} there is a residual chiral U$(1)_{G_2}$ symmetry generated by the $G_2=T_3+\frac{T_8}{\sqrt{3}}+\frac{\textrm{PQ}}{3}$ generator which remains unbroken after the $\rho$ and $\chi$ fields acquire vevs. As it happens, the U$(1)_{G_2}$ symmetry is also chiral in the quarks, leaving three of them massless. Therefore, the accidental U$(1)_{G_2}$ has to be broken in order to generate mass terms for the massless up- and down-type quarks, which can be achieved by introducing some extra fields. The consequences of this kind of accidental chiral symmetry were studied for the first time in the economical 3-3-1 model in Ref. \cite{PhysRevD.91.037302}.  In the original 3-3-1 models, the U$(1)_{G_2}$ symmetry was not explicitly recognised but its breakdown, avoiding massless fermions, was done through an enlarged set of nontrivial SU$(3)_{L}$ multiplets of scalar fields, e.g. three triplets~\cite{Singer:1980sw,Frampton:1992wt,Montero:1992jk,Pleitez:1992xh,Foot:1994ym,Pleitez:1994pu,Ozer:1995xi,Hoang:1995vq} and three triplets plus a sextet~\cite{Foot:1992rh}. Other studies have tried to address this issue, while assuming only two triplets in the scalar sector, via radiative corrections~\cite{Ponce:2002sg} or non-renormalisable operators~\cite{Ferreira:2011hm,Dong:2014esa}. Nevertheless, none of these constructions took scale invariance into account.

\begin{table}[ht]
\begin{tabular}{|c|c|c|c|c|c|c|c|c|}
\hline 
 & \, $Q_{1 L}$  & \, $Q_{aL}$  & \, ($u_{iR}$, $U_R$)  & \,($d_{iR}$, $D_{aR}$)  & \, $F_{iL}$  & \, $e_{iR}$  & \, $\rho$  & \, $\chi$ \tabularnewline
\hline 
\hline 
$\textrm{U\ensuremath{\left(1\right)}}_{X}$ & $1/3$ & $0$    & $2/3$  & $-1/3$  & $-1/3$  & $-1$  & $2/3$  & $-1/3$ \tabularnewline
\hline 
$\textrm{U\ensuremath{\left(1\right)}}_{B}$ & $1/3$ & $1/3$  & $1/3$  & $1/3$  & $0$  & $0$  & $0$  & $0$ \tabularnewline
\hline 
$\textrm{U\ensuremath{\left(1\right)}}_{\textrm{PQ}}$  & $1$ & $-1$ & $0$  & $0$  & $-1/2$  & $-3/2$  & $1$  & $1$ \tabularnewline
\hline 
\end{tabular}
\caption{U$\left(1\right)$ charges in the model with only two scalar triplets.
}
\label{table1pq}
\end{table}

As shown in Table \ref{table1pq}, in addition to invariance under the gauge group $\text{U}(1)_X$ and the accidental chiral symmetry $\text{U}(1)_{\text{PQ}}$, the model with only two triplets is also invariant under $\text{U}(1)_B$, reflecting the conservation of Baryon number. However, it is not invariant under a Lepton number symmetry, $\text{U}(1)_L$. Therefore, it is natural to investigate whether a $\text{U}(1)_L$, or $\text{U}(1)_{B-L}$, symmetry can be easily accommodated in the model.

\subsection{$B-L$ as a residual gauge symmetry}

Perhaps, the most straightforward way we can define $\text{U}(1)_L$ is by assigning a $+1$ charge to all lepton multiplets, $F_{iL}$ and $e_{iR}$, and vanishing charges to the remaining fields. In this case, one can easily check that the last term in Eq. (\ref{yuklel}) breaks $\text{U}(1)_L$ explicitly. Thus, once such a term is removed from the Lagrangian, the $\text{U}(1)_L$ symmetry is realised. This Abelian symmetry is, however, anomalous and can not be promoted to local; the same applies to the usual $\text{U}(1)_{B-L}$, Baryon number minus Lepton number symmetry.

An alternative realisation of a Lepton number symmetry is obtained by allowing the third components of the fermion triplets, which contain non-SM fermions, as well as their right-handed partners, to carry non-conventional Lepton number charges. This can be achieved with a new $\text{U}(1)$ symmetry plus the help of the $T_8=(2\sqrt{3})^{-1} \times\mbox{diag}(1,1,-2)$ generator of $\text{SU}(3)_L$, which tells the third component of (anti-)triplets from the first two~\cite{Dong:2013wca,Dong:2015yra}. To keep the discussion concise, let us define the $B-L$ symmetry, the difference between the $\text{U}(1)_{B}$ symmetry, shown in Table \ref{table1pq}, and our Lepton number symmetry $L$, as 
\be \label{BmL}
B-L = \beta^\prime T_8 + N\bf{1}_{3\times 3},
\ee
where different $\beta^\prime$ values define different $B-L$ (or $L$) symmetries, and $N$ represents the charge of a given field under the new Abelian symmetry $\text{U}(1)_{N}$. Notice that the $B-L$ generator resembles the electric charge generator, $Q$, defined in Eq. (\ref{Q}). In fact, just like $Q$ emerges from the spontaneous breaking of $\text{SU}(3)_L$ and $\text{U}(1)_X$ as a residual symmetry, $B-L$ is a residual symmetry from the breaking of the $\text{SU}(3)_L$ and $\text{U}(1)_N$ symmetries, as we show next.

In order to fix a value for $\beta^\prime$, we conveniently impose the $B-L$ charge of the third component of $F_{iL}$, {\it i.e.} $(N_{iR})^c$, to be $+1$:\footnote{Notice that the $N_{iR}$ form Dirac mass terms with the left-handed neutrinos $\nu_{iL}$, via Eq. (\ref{yuklel}); thus we are assuming that the $N_{iR}$ fields -- effectively right-handed neutrinos -- carry the same $B-L$ number as $\nu_{iL}$.}
\be
(B-L) F_{iL} = \mbox{diag}\left(\frac{\beta^\prime}{2\sqrt{3}}+ N_{F_{L}},\frac{\beta^\prime}{2\sqrt{3}}+ N_{F_{L}},-\frac{\beta^\prime}{\sqrt{3}} + N_{F_{L}}\right) F_{iL} \equiv \mbox{diag}(-1,-1,+1) F_{iL},
\ee 
which translates into $\beta^\prime= -4/\sqrt{3}$ and $N_{F_{L}} =- 1/3$. With $\beta^\prime$ fixed, we can derive the $B-L$ charges of all the other fermions simply by imposing that the SM leptons and quarks carry charges equal to $-1$ and $1/3$, respectively. Moreover, we can extend the symmetry to the scalar sector by taking into account the Yukawa interactions. For example, from Eq. (\ref{yuklel}), we derive the $B-L$ charge of $\rho$, whereas from the last term in Eq. (\ref{yukq}) we get the $B-L$ charge of $\chi$. Thus, we obtain the set of $N$ and $B-L$ charges of scalars and fermions shown in Table \ref{tableBL}.

\begin{table}[th]
\tabcolsep=0.03cm
\begin{tabular}{|c|c|c|c|c|c|c|c|c|c|c|}
\hline 
 & \, $Q_{1 L}$ \, & \, $Q_{aL}$ \, & \, $u_{i R}$\, & \, $U_R$ \, & \,$d_{i R} $ \, & $D_{aR}$ \, & \, $F_{i L}$ \, & \, $e_{i R}$ \, & \, $\rho$ \, & \, $\chi$ \,\tabularnewline
\hline 
$N$  & $1$  & $-1/3$  & $1/3$  & $7/3$  & $1/3$  & $-5/3$  & $-1/3$  & $-1$ & $2/3$ & $-4/3$ \tabularnewline
\hline 
\,$B-L$\, &\, $(1/3,\,1/3,\,7/3)^T\,$  &\, $(1/3,\,1/3,\,-5/3)^T\,$  & $1/3$  & $7/3$  & $1/3$  &\, $-5/3$\, &\, $(-1,\,-1,\,1)^T\,$  & $-1$ & $\,(0,\,0,\,2)^T\,$ & $\,(-2,\,-2,\,0)^T\,$ \tabularnewline
\hline
\end{tabular}
\caption{ $\text{U}(1)_N$ and $\text{U}(1)_{B-L}$ charges in the model with only two scalar triplets. \label{tableBL}}
\end{table}

From Table \ref{tableBL}, we can deduce that, despite both $\text{SU}(3)_L$ and $\text{U}(1)_N$ being spontaneously broken by $\langle \rho\rangle=(0,v_\rho/\sqrt{2},0)^T$ and $\langle \chi\rangle=(0,0,v_\chi/\sqrt{2})^T$, the combination that defines $B-L$ remains fully conserved so that $B-L$ in Eq. (\ref{BmL}), similar to $Q$ in Eq. (\ref{Q}), represents a residual symmetry. Furthermore, the new Abelian group $\text{U}(1)_N$ is anomaly-free and so is $B-L$ and, as such, can be promoted to local. Then, the gauge symmetry of the model is $\text{SU}(3)_C\otimes \text{SU}(3)_L\otimes \text{U}(1)_X\otimes \text{U}(1)_N$. Finally, we point out that by imposing $\text{U}(1)_N$, or equivalently $B-L$, as given in Table \ref{tableBL}, the terms in Eq. (\ref{yukq}) governed by the Yukawa couplings $Y^D_{1a}, Y^U_{a}, Y^u_{1i}, Y^d_{ai}$ are forbidden. The absence of such terms does not culminate in new massless quarks in the spectrum but rather helps to turn off any mixing between the SM and the exotic quarks.

\subsection{The scale-invariant 3-3-1-1 model}

In this subsection, we propose an extension to tackle the main drawbacks of the model discussed in the previous sections. In order to break the residual chiral symmetry, $\text{U}(1)_{\text{PQ}}$, that renders some of the SM fermions massless, we follow the approach adopted in Ref.~\cite{Dias:2020ryz} and introduce heavy vector-like fermions. Then, the previously massless SM fermions become massive via a seesaw-type mechanism, mediated by the new fermions. The scalar sector is minimally enlarged by a complex and a real 3-3-1 gauge singlet. As we will see in the following sections, the scalar singlets play important roles in the consistency of the dynamical symmetry breaking mechanism, generating heavy masses for the $B-L$ vector boson and the new seesaw-mediating fermions.

In the quark sector, we consider two additional quark representations 
\begin{eqnarray}\label{newq}
{\cal{Q}}_{L,R}=({\cal{U}}_{1},\,q^{(5/3)},\,{\cal{U}}_2)^T_{L,R} \,&& \sim(\mathbf{3},\mathbf{\bar{3}},1, 5/3), \label{qextra1}   \\
{\cal{Q}}_{b\,L,R}=({\cal{D}}_{b},\,q^{(-4/3)}_b,\,{\cal{D}'}_{b})^T_{L,R} \,&& \sim(\mathbf{3},\mathbf{3},-2/3,-1), \label{qextra2}
\end{eqnarray}
with $b=1,2$ and the numbers in parentheses representing the field's transformations under the $\text{SU}(3)_C\otimes \text{SU}(3)_L\otimes \text{U}(1)_X\otimes \text{U}(1)_N$ gauge group. The first and third components of ${\cal{Q}}_{L,R}$ (${\cal{Q}}_{b\,L,R}$) carry the same electric charge as the up-type (down-type) quarks, while the second component has an exotic electric charge of $+5/3$ ($-4/3$). Notice also that since the new quarks are introduced as vector-like pairs, their contributions to anomaly coefficients vanish identically.

We also introduce the following neutral leptons to ensure a viable neutrino spectrum
\be\label{newl}
{S}_{i\,L,R}\sim(\mathbf{1},\mathbf{1},0, 1), 
\ee 
and, two scalar fields belonging the following representations  
\bea\label{news}
\sigma \sim (\mathbf{1},\mathbf{1},0,2),\quad \varphi \sim(\mathbf{1},\mathbf{1},0,0), 
\eea 
in which $\varphi$ is real.

The choices of the new fields in Eqs. \eqref{qextra1}-\eqref{news} are guided by the need to generate masses for the SM fermions and by our aspiration to maintain the scalar sector as minimal as possible. Consider, for instance, the up-type quark sector, where one quark ($u_{1}$) is left massless in the model without extra fields. To generate a mass for this field, we need a term of the form $m_u \,\overline{u_{1L}}\,{\cal{U}}_R$, where the ${\cal{U}}_R$ is an extra up-type quark and $ u_{1L}	\subset Q_{1L}$. Since we prefer to avoid introducing new scalar triplets, the new quark ${\cal{U}}_R$ cannot be a singlet of $\text{SU}(3)_L$. Thus, the next simpler possibility is that it belongs to the $\mathbf{3}$ or $\mathbf{\bar{3}}$ representation of SU$(3)_L$. Amongst the alternatives, we select ${\cal{Q}}_R$, defined in Eq. (\ref{qextra1}), which contributes to the generation of quark masses via the SU$(3)_L$ anti-symmetric operator $\overline{Q_{1L}}{\cal{Q}}_R\rho^{*}$. Moreover, its left-handed partner: ${\cal{Q}}_L$ must be added to cancel anomalies. Lastly, to make sure that all quarks in ${\cal{Q}}_{L,R}$ also become massive following the dynamical breaking of scale invariance, we add the real scalar singlet $\varphi$ in Eq. (\ref{news}) and, consequently, the operator $\overline{\cal{Q}}_{L} {\cal{Q}}_{R}\varphi$. In conclusion, with the introduction of the quark triplets: ${\cal{Q}}_{L,R}$ and the scalar singlet: $\varphi$, the Yukawa Lagrangian is enlarged in such a way that not only all the up-type but also the new quarks become massive\footnote{See the details on the mass generation mechanism in the Sec. \ref{subsecQuarks}}. Similar reasoning is used to introduce the new quarks ${\cal{Q}}_{bL,R}$, defined in Eq. (\ref{qextra2}), which contribute to generating masses to all down-type quarks.

Finally, the scalar $\sigma$ plays important roles in the breaking of the $B-L$ symmetry and mass generation for the corresponding vector boson, as well as in the neutrino sector, where -- together with the new neutral leptons ${S}_{i\,L,R}$ in Eq. (\ref{newl}) -- it contributes to neutrino mass generation through the terms  $ \sigma^*\overline{(S_{i R})^{C}} S_{j R}$ and  $\sigma \overline{S_{i L}} (S_{j L})^{C}$, as we show in Sec. \ref{leptonsector}.

\section{Scalar Sector}\label{scsect}

With the assumption of scale invariance, the spontaneous breakdown of gauge symmetries can be triggered dynamically through the Coleman-Weinberg mechanism~\cite{Coleman:1973jx,Weinberg:1973am}. We proceed to present a detailed study of such a mechanism at the one-loop level. In order to do that, we make use of the Gildener-Weinberg method~\cite{Gildener:1976ih}, which allows calculating the effective potential perturbatively  on the flat directions defined in scenarios with multiple scalar fields. 

Let us start with the most general renormalisable scale-invariant scalar potential respecting the gauge symmetry of the model: 
\begin{eqnarray}
V_{0}&=&\lambda_{\chi}\left(\chi^{\dagger}\chi\right)^{2}+\lambda_{\rho}\left(\rho^{\dagger}\rho\right)^{2}+\lambda_{\chi\rho}\left(\chi^{\dagger}\chi\right)\left(\rho^{\dagger}\rho\right)+\lambda_{\chi\rho}^\prime\left(\chi^{\dagger}\rho\right)\left(\rho^{\dagger}\chi\right)  \nonumber \\
&+&\lambda_{\chi\varphi}\left(\chi^{\dagger}\chi\right) \varphi^2+ \lambda_{\chi\sigma}\left(\chi^{\dagger}\chi\right)\left(\sigma^{*}\sigma\right) + \lambda_{\rho\varphi}\left(\rho^{\dagger}\rho\right)\varphi^2 + \lambda_{\rho\sigma} \left(\rho^{\dagger}\rho\right) \left(\sigma^{*}\sigma\right)\nonumber\\
&+& \lambda_{\varphi}\varphi^4 + \lambda_{\sigma}\left(\sigma^{*}\sigma\right)^{2}+\lambda_{\varphi\sigma}\left(\sigma^{*}\sigma\right)\varphi^2, \label{treelevelpotential}
\end{eqnarray}
where the two scalar triplets,  $\rho$ and $\chi$, the complex singlet, $\sigma$, and the real singlet, $\varphi$, are written as 
\bea\label{scalars}
\rho^T&=&\left(\rho^{+}_{1},\, \frac{S_\rho + i A_\rho}{\sqrt{2}},\,\rho^{+}_{3}\right); \quad \quad \quad \, \sigma= \frac{S_\sigma+ i A_\sigma}{\sqrt{2}};  \\
\chi^T&=&\left(\frac{S_{\chi_1}+iA_{\chi_1}}{\sqrt{2}},\,\chi_2^{-}, \,\frac{S_{\chi_3} + i A_{\chi_3}}{\sqrt{2}} \right); \quad \varphi= \frac{S_\varphi}{\sqrt{2}}.
\eea
Before considering the one-loop effective potential, we determine the conditions  for the tree-level potential to be bounded from below. The constraints on the scalar potential couplings coming from this are straightforwardly obtained noting that $V_0$ in Eq. (\ref{treelevelpotential}) can be written as $V_0={\bf h}^T\mathbf{\Lambda}(\theta){\bf h}$, where ${\bf h}\equiv(|\chi|^{2},|\rho|^{2},|\sigma^{2}|,|\varphi|^{2})^T\geq0$, and $\mathbf{\Lambda}(\theta)$ is the matrix 
\begin{eqnarray}
\mathbf{\Lambda}(\theta)=\left(
\begin{array}{cccc}
 \lambda_\chi  & \frac{ \lambda_{\chi \rho} +\lambda'_{\chi \rho}
    \theta}{2} &
   \frac{\lambda_{\chi \sigma}}{2} & \frac{\lambda_{\chi
   \varphi} }{2} \\
 \frac{\lambda_{\chi \rho} +\lambda'_{\chi \rho}
    \theta }{2} & \lambda_\rho  & \frac{\lambda_{\rho \sigma}}{2} & \frac{\lambda_{\rho \varphi} }{2} \\
 \frac{\lambda_{\chi \sigma}}{2} & \frac{\lambda_{\rho \sigma}}{2} & \lambda_{\sigma}  & \frac{\lambda_{\varphi \sigma}}{2} \\
 \frac{\lambda_{\chi \varphi}}{2} & \frac{\lambda_{\rho \varphi}}{2} & \frac{\lambda_{\varphi \sigma}}{2} &
   \lambda_\varphi \\
\end{array}
\right),
\label{lambda}
\end{eqnarray}
where $0\leq \theta \leq1$ is the orbit parameter defined as $\theta=\hat{\chi}^{*}_i\hat{\rho}_i\hat{\rho}^{*}_j\hat{\chi}_j$, with $i,j=1,2,3$, and $\hat{\chi}_i,\,\hat{\rho}_i=\chi_i/|\chi|,\,\rho_i/|\rho|$.

Thus, $V_0$ is bounded from below if it is positive for all possible values of $\bf {h}$. As $\bf {h}\geq0$, this happens  if $\mathbf{\Lambda}(\theta)$ is copositive \cite{Kannike:2012pe}. Using the Cottle-Habetler-Lemke theorem \cite{
COTTLE1970295} and noting that $\theta$ can take two values, $\theta=0$ for $\lambda'_{\chi \rho}>0$ and $\theta=1$, otherwise, we obtain the following conditions that make the $\mathbf{\Lambda}(\theta=0,1)$ copositive
\bea \label{cop_1}
&&\lambda_{\chi} \geq 0,\quad\quad \lambda_{\rho} \geq  0, \quad\quad \lambda_{\sigma} \geq 0,\quad\quad \lambda_{\varphi} \geq 0, \nonumber \\
&&\overline{\lambda}_1 \equiv 2 \sqrt{\lambda_\rho \lambda_\chi }+\overline{\lambda}_{\chi \rho} \geq 0, \quad \overline{\lambda}_2 \equiv 2\sqrt{\lambda_\sigma  \lambda_\chi }+\lambda_{\chi \sigma} \geq 0, \quad \overline{\lambda}_3 \equiv 2 \sqrt{\lambda_\rho  \lambda_\sigma}+\lambda_{\rho \sigma} \geq 0, \nonumber \\
&& \overline{\lambda}_4 \equiv 2 \sqrt{\lambda_\varphi  \lambda_\chi}+\lambda_ {\chi \varphi} \geq 0, \quad \overline{\lambda}_5 \equiv 2 \sqrt{\lambda_\sigma  \lambda_\varphi }+\lambda_{\varphi \sigma} \geq 0, \quad \overline{\lambda}_6 \equiv  2 \sqrt{\lambda_\rho  \lambda_\varphi }+\lambda_{\rho \varphi} \geq 0, \nonumber \\
&& \sqrt{\overline{\lambda}_1 \overline{\lambda}_2 \overline{\lambda}_3 }+2 \sqrt{\lambda_\rho  \lambda_\sigma  \lambda_\chi }+\overline{\lambda}_{\chi \rho} \sqrt{\lambda_\sigma } +\lambda_{\chi \sigma} \sqrt{\lambda_\rho} +\lambda_{\rho \sigma} \sqrt{\lambda_\chi}\geq 0, \nonumber\\
&&\sqrt{\overline{\lambda}_2 \overline{\lambda}_4 \overline{\lambda}_5}+2 \sqrt{\lambda_\sigma  \lambda_\varphi  \lambda_\chi }+\lambda_{\chi \varphi}\sqrt{\lambda_\sigma} +\lambda_{\chi \sigma}\sqrt{\lambda_\varphi}+\lambda_{\varphi \sigma}  \sqrt{\lambda_\chi }\geq 0, \nonumber\\
&&\sqrt{\overline{\lambda}_1 \overline{\lambda}_4 \overline{\lambda}_6}+2\sqrt{\lambda_\rho  \lambda_\varphi \lambda_\chi }+\overline{\lambda}_{\chi \rho} \sqrt{\lambda_\varphi }+ \lambda_{\chi \varphi}\sqrt{\lambda_\rho}+\lambda_{\rho \varphi} \sqrt{\lambda_\chi }\geq 0, \nonumber\\
&&\sqrt{\overline{\lambda}_3 \overline{\lambda}_5 \overline{\lambda}_6}+2 \sqrt{\lambda_\rho  \lambda_\sigma \lambda_\varphi }+\lambda_{\varphi \sigma} \sqrt{\lambda_\rho}+\lambda_{\rho \sigma}  \sqrt{\lambda_\varphi}+\lambda_{\rho \varphi}  \sqrt{\lambda_\sigma} \geq 0, \nonumber \\
&&\det[\mathbf{\Lambda}(\theta=0,1)]\geq 0,
\eea
where $\overline{\lambda}_{\chi \rho}$ takes two values: $\lambda_{\chi \rho}$ (when $\theta =0$) and $\lambda_{\chi \rho}+ \lambda'_{\chi \rho}$ (when $\theta =1$). 
To implement the Gildener-Weinberg method, we need to find the flat direction, {\it i.e.}, a direction in the field space along which the potential and its first derivative vanish simultaneously at tree level. In other words, the direction in the vacuum surface, $\mathbf{N}=\mathbf{n}$, which satisfies: 
$i)$  $\nabla_{\mathbf{N}}V_0(\mathbf{N})|_{\mathbf{N}=\mathbf{n}}=0$ 
and $ii)$  $V_0(\mathbf{n})=0$.
From the condition $i)$, we find
\begin{eqnarray}\label{DofV}
\mathbf{\Lambda}_0.\mathbf{n}^2=
\left(
\begin{array}{cccc}
 \lambda_\chi  & \frac{\lambda_{\chi \rho} }{2} &
   \frac{\lambda_{\chi \sigma}}{2} & \frac{\lambda_{\chi \varphi}}{2} \\
 \frac{\lambda_{\chi \rho} }{2} & \lambda_{\rho}  &
   \frac{\lambda_{\rho \sigma}}{2} & \frac{\lambda_{\rho \varphi} }{2} \\
 \frac{\lambda_{\chi \sigma}}{2} & \frac{\lambda_{\rho \sigma}}{2} & \lambda_\sigma  & \frac{\lambda_{\varphi \sigma}}{2} \\
 \frac{\lambda_{\chi \varphi}}{2} & \frac{\lambda_{\rho \varphi}}{2} & \frac{\lambda_{\varphi \sigma}}{2} &
   \lambda_{\varphi} \\
\end{array}
\right)
\left(
\begin{array}{c}
n_{\chi}^2\\
n_{\rho}^2\\
n_{\sigma}^2 \\
n_{\varphi}^2
\end{array}
\right)
=\left(
\begin{array}{c}
0\\
0\\
0\\
0\end{array}
\right),
\end{eqnarray}
where ${\mathbf{n}}^T=(n_\chi, n_\rho, n_\sigma, n_\varphi)$ is the unit vector in the scalar field space evaluated in the vacuum. We also have that $\mathbf{n}^2$ stands for $(n_\chi^2,n_\rho^2, n_\sigma^2,n_\varphi^2)^T$, and $\mathbf{\Lambda}_0$ is equal to the quartic coupling matrix given in Eq. (\ref{lambda}) with $\theta=0$, {\it i.e.} $\mathbf{\Lambda}(\theta=0)$. 

In order to obtain non-trivial solutions for $\mathbf{n}^2$ in Eq. \eqref{DofV}, it is necessary that $\det \mathbf{\Lambda}_0=0$ \cite{Kannike:2019upf}. This condition can be satisfied if for a given renormalisation scale, $\mu_{0}$, the $\lambda_{\sigma}$ coupling assumes the value  \begin{eqnarray}
\lambda_{\sigma}|_{\mu_{0}}&=&\frac{-4 \lambda_\rho  \left(\lambda_\varphi  \lambda_{\chi \sigma}^2+\lambda_{\varphi \sigma}^2 \lambda_\chi 
-\lambda_{\varphi \sigma} \lambda_{\chi \sigma}  \lambda_{\chi \varphi}\right)+\lambda_{\rho \sigma}^2 \left(\lambda_{\chi \varphi}^2-4 \lambda_\varphi  \lambda_{\chi} \right)+(\lambda_{\varphi \sigma} \lambda_{\chi \rho} -\lambda_{\rho \varphi} \lambda_{\chi \sigma})^2}
  {4 \left(\lambda_\rho\left(\lambda_{\chi \varphi}^2-4 \lambda_\varphi  \lambda_\chi \right)+\lambda_{\rho \varphi}^2 \lambda_\chi -\lambda_{\rho \varphi} \lambda_{\chi \rho}  \lambda_{\chi \varphi} +\lambda_\varphi  \lambda_{\chi \rho}^2\right)} \nonumber  \\
&+&\frac{\lambda_{\rho\sigma} (4\lambda_{\rho \varphi} \lambda_{\varphi \sigma} 
 \lambda_\chi -2 \lambda_{\chi \varphi} (\lambda_{\rho
   \varphi}  \lambda_{\chi \sigma} +\lambda_{\varphi \sigma} 
   \lambda_{\chi \rho})+4 \lambda_\varphi  \lambda_{\chi \rho} \lambda_{\chi \sigma})}{4 \left(\lambda_\rho  
   \left(\lambda_{\chi \varphi}^2-4 \lambda_\varphi  \lambda_\chi \right)+\lambda_{\rho \varphi}^2 \lambda_\chi -\lambda_{\rho \varphi} \lambda_{\chi \rho}  \lambda_{\chi \varphi} +\lambda_\varphi  \lambda_{\chi \rho}^2\right)} .
\label{lambdasigma}
\end{eqnarray}
Once this condition is satisfied, we obtain the following solutions for $\mathbf{n}^2$ in Eq. (\ref{DofV}) 
\begin{eqnarray}
n_{\chi}^{2}&=&\frac{\lambda_{\varphi \sigma} (-2 \lambda_{\rho}  (\lambda_{\chi \sigma} +\lambda_{\chi \varphi})+\lambda_{\rho \sigma} (-2 \lambda_{\rho \varphi}+\lambda_{\chi \rho}+\lambda_{\chi \varphi})+\lambda_{\rho \varphi} (\lambda_{\chi \rho} +\lambda_{\chi
   \sigma}))}{2\,\textrm{den}} \nonumber   \label{nsquared11} \\
&&+\frac{2 \lambda_\sigma \left(-4 \lambda_\rho 
   \lambda_\varphi +2 \lambda_\rho  \lambda_{\chi \varphi}
   +\lambda_{\rho \varphi}^2-\lambda_{\rho \varphi} (\lambda_{\chi \rho}+\lambda_{\chi \varphi} )+2 \lambda_{\varphi} 
   \lambda_{\chi \rho}\right)+\lambda_{\varphi \sigma}^2 (2
   \lambda_\rho -\lambda_{\chi \rho} )}{2\,\textrm{den}} \nonumber \\   
   &&+\frac{2 \lambda_\varphi  \left(2 \lambda_\rho  \lambda_{\chi \sigma}+\lambda_{\rho \sigma}^2-\lambda_{\rho \sigma} (\lambda_{\chi \rho}+\lambda_{\chi \sigma})\right)-(\lambda_{\rho \sigma}
   -\lambda_{\rho \varphi} ) (\lambda_{\rho \sigma } \lambda_{\chi \varphi} -\lambda _{\rho \varphi} \lambda_{\chi \sigma})}{2\,\textrm{den}}; \label{nchi} \\   
   n_{\rho}^2&=&\frac{2 \lambda_\chi  \left(-\lambda_{\varphi \sigma} (\lambda_{\rho
   \sigma}+\lambda_{\rho \varphi})+2 \lambda_{\rho \sigma} 
   \lambda_\varphi +2 \lambda_\sigma  (\lambda_{\rho \varphi}
   -2 \lambda_\varphi )+\lambda_{\varphi \sigma}^2\right)+\lambda_{\rho \sigma} 
   \lambda_{\chi \sigma}  \lambda_{\chi \varphi}-2 \lambda_{\rho \varphi }
   \lambda_{\sigma } \lambda_{\chi \varphi}}{2\,\textrm{den}} \nonumber \\
    &&+\frac{\lambda_{\varphi \sigma} (\lambda_{\chi \varphi }
   (\lambda_{\rho \sigma}+\lambda_{\chi \rho} -2\lambda_{\chi \sigma} )+\lambda_{\rho \varphi} \lambda_{\chi \sigma}
   +\lambda_{\chi \rho} \lambda_{\chi \sigma})  -\lambda_{\varphi
   \sigma} ^2 \lambda_{\chi \rho}-\lambda_{\rho
   \sigma}  \lambda_{\chi \varphi}^2-\lambda_{\rho
   \varphi}  \lambda_{\chi \sigma} ^2}{2\,\textrm{den}}\nonumber \\
&&+\frac{2 \lambda_{
   \varphi}  \left(-\lambda_{\chi \sigma } (\lambda_{\rho \sigma}
   +\lambda_{\chi \rho})+2 \lambda_{\sigma} \lambda_{\chi \rho}
   +\lambda_{\chi \sigma}^2\right) +\lambda_{\rho \varphi} 
   \lambda_{\chi \sigma} \lambda _{\chi \varphi}-2 \lambda_{\sigma } \lambda_{\chi \rho}  \lambda_{\chi \varphi} +2
   \lambda_{\sigma}  \lambda_{\chi \varphi}^2}{2\,\textrm{den}}; \label{nrho} \\
  n_{\sigma}^2&=&\frac{2 \lambda_\chi  \left(-4 \lambda_\rho  \lambda_\varphi +2
   \lambda_\rho  \lambda_{\varphi \sigma} -\lambda_{\rho \sigma}
    \lambda_{\rho \varphi} +2 \lambda _{\rho \sigma } \lambda_\varphi +\lambda_{\rho \varphi} ^2-\lambda_{\rho \varphi} 
   \lambda_{\varphi \sigma} \right)+4 \lambda_\rho  \lambda_ \varphi  \lambda_{\chi \sigma} -2 \lambda_\rho  \lambda_{\varphi \sigma}  \lambda_{\chi \varphi}}{2\,\textrm{den}}\nonumber \\
   && +\frac{2 \lambda_\varphi 
   \lambda_{\chi \rho}  (-\lambda_{\rho \sigma}+\lambda_{\chi
   \rho} -\lambda_{\chi \sigma})+\lambda_{\rho \sigma}  \lambda_{
   \chi \rho}  \lambda_{\chi \varphi} -\lambda_{\rho \sigma} 
   \lambda_{\chi \varphi} ^2+\lambda_{\rho \varphi} ^2 (-\lambda_{\chi \sigma})-2 \lambda_\rho 
   \lambda_{\chi \sigma}  \lambda_{\chi \varphi}+2 \lambda_{\rho}
    \lambda_{\chi \varphi} ^2 }{2\,\textrm{den}} \nonumber \\ 
   &&+\frac{\lambda_{\varphi \sigma}  \lambda _{\chi \rho} 
   (\lambda_{\rho \varphi} -\lambda_{\chi \rho} +\lambda_{\chi
   \varphi} )+\lambda_{\rho \varphi}  \lambda_{\chi \rho} 
   \lambda_{\chi \sigma} -2 \lambda_{\rho \varphi} \lambda_{\chi
   \rho} \lambda_{\chi \varphi }+\lambda_{\rho \varphi}  \lambda_{\chi \sigma} \lambda_{\chi \varphi} +\lambda_{\rho \sigma}  \lambda_{\rho \varphi} \lambda_{\chi \varphi}}{2\,\textrm{den}};\label{nsigma} \\ 
    n_{\varphi}^2&=&\frac{2 \lambda_\chi  \left(-4 \lambda_\rho  \lambda_\sigma +2
   \lambda_\rho  \lambda_{\varphi \sigma} +\lambda_{\rho \sigma}
   ^2-\lambda_{\rho \sigma} (\lambda_{\rho \varphi} +\lambda_{\varphi \sigma})+2 \lambda_{\rho \varphi } \lambda_{\sigma}
   \right)+\lambda_{\rho \varphi }
   \lambda_{\chi \rho} \lambda_{\chi \sigma} -\lambda_{\rho
   \varphi } \lambda_{\chi \sigma}^2}{2\,\textrm{den}} \nonumber  \\ 
   &&+\frac{4 \lambda_{\rho}  \lambda_{\sigma} \lambda_{\chi
   \varphi} -2 \lambda_{\rho}  \lambda_{\varphi \sigma} \lambda_{\chi \sigma} +2 \lambda _{\rho}  \lambda_{\chi \sigma}^2-2
   \lambda_\rho  \lambda_{\chi \sigma}  \lambda_{\chi \varphi}
   -\lambda_{\rho \sigma}^2 \lambda_{\chi \varphi }-\lambda_{\varphi \sigma} 
   \lambda_{\chi \rho} ^2+\lambda _{\varphi \sigma}  \lambda_{\chi
   \rho } \lambda_{\chi \sigma}}{2\,\textrm{den}} \nonumber \\
   &&+\frac{\lambda_{\rho \sigma}  (\lambda_{\rho \varphi } \lambda_{\chi \sigma}
   +\lambda_{\varphi \sigma}  \lambda_{\chi \rho} +\lambda_{\chi
   \varphi } (\lambda_{\chi \rho} +\lambda_{\chi \sigma} )-2
   \lambda_{\chi \rho}  \lambda_{\chi \sigma} )+2 \lambda_{\sigma}
    \lambda_{\chi \rho}  (-\lambda_{\rho \varphi} +\lambda_{\chi
   \rho} -\lambda_{\chi \varphi} )}{2\,\textrm{den}},
   \label{nsquared}
\end{eqnarray}
where $\textrm{den}$ is defined as
\begin{eqnarray}
\textrm{den}\equiv&&\lambda_\varphi  \left(4 \lambda _\sigma  (\lambda _{\chi \rho}
   -\lambda_\rho )+4 \lambda_\rho  \lambda_{\chi \sigma}
   +\lambda_{\rho \sigma}^2-2 \lambda_{\rho \sigma } (\lambda_{\chi \rho} +\lambda_{\chi \sigma} )+(\lambda_{\chi \rho}
   -\lambda_{\chi \sigma})^2\right) \nonumber \\
   &&+\lambda_\chi  \left(4 \lambda_\varphi  (\lambda_{\rho \sigma}
   -\lambda_{\rho})-4 \lambda _{\sigma}(\lambda_{\rho} -\lambda_{\rho \varphi} +\lambda_{\varphi} )+4 \lambda_{\rho}  \lambda_{\varphi \sigma} -2 \lambda_{\varphi \sigma}  (\lambda_{\rho
   \sigma}+\lambda_{\rho \varphi})+(\lambda_{\rho \sigma}
   -\lambda_{\rho \varphi} )^2+\lambda_{\varphi \sigma}
   ^2\right) \nonumber \\
   && \lambda_{\varphi \sigma} \left(-2 \lambda_\rho  (\lambda_{\chi
   \sigma} +\lambda_{\chi \varphi})+\lambda_{\chi \rho} 
   (\lambda_{\rho \sigma} +\lambda_{\rho \varphi} +\lambda_{\chi
   \sigma}+\lambda_{\chi \varphi} )-(\lambda_{\rho \sigma}
   -\lambda_{\chi \sigma}) (\lambda_{\rho \varphi} -\lambda_{ \chi \varphi} )-\lambda_{\chi \rho} ^2\right)+4 \lambda_{\rho}
    \lambda_{\sigma}  \lambda_{\chi \varphi}\nonumber \\
   && \lambda_{\varphi \sigma}^2 (\lambda_{\rho} -\lambda_{\chi \rho})+\lambda_{\rho}  \lambda_{\chi \sigma}^2-2 \lambda_{\rho}
   \lambda_{\chi \sigma} \lambda_{\chi \varphi}+\lambda_{\rho} 
   \lambda_{\chi \varphi}^2-\lambda_{\rho \sigma}^2\lambda_{\chi \varphi}+\lambda_{\rho \sigma} \lambda_{\rho \varphi} 
   \lambda_{\chi \sigma}+\lambda_{\rho \sigma} \lambda_{\rho
   \varphi} \lambda_{\chi \varphi}-2
   \lambda_{\sigma } \lambda_{\chi \rho}  \lambda_{\chi \varphi} \nonumber \\
   &&-\lambda_{\rho \sigma}  \lambda_{\chi \rho}  \lambda_{\chi \sigma}
   +\lambda_{\rho \sigma} \lambda_{\chi \rho } \lambda_{\chi
   \varphi}+\lambda_{\rho \sigma } \lambda_{\chi \sigma} 
   \lambda_{\chi \varphi} -\lambda_{\rho \sigma } \lambda_{\chi
   \varphi}^2+\lambda_{\rho \varphi} ^2 \lambda_{\sigma}
   -\lambda_{\rho \varphi}^2 \lambda_{\chi \sigma}-2 \lambda
   _{\rho \varphi}  \lambda_{\sigma } \lambda_{\chi \rho}+\lambda_{\sigma } \lambda_{\chi \varphi}^2 \nonumber  \\
   &&-2\lambda_{\rho \varphi } \lambda_{\sigma}  \lambda_{\chi \varphi}
   +\lambda_{\rho \varphi}  \lambda_{\chi \rho} \lambda_{\chi
   \sigma}-\lambda_{\rho \varphi } \lambda_{\chi \rho}  \lambda_{
   \chi \varphi}-\lambda_{\rho \varphi} \lambda_{\chi \sigma}
   ^2+\lambda_{\rho \varphi } \lambda_{\chi \sigma} \lambda
 _{\chi \varphi }+\lambda_{\sigma } \lambda_{\chi \rho}^2.
\end{eqnarray}
The condition $ii)$ for the flat direction,  $V_0(\mathbf{n})=0$, is also satisfied, and, due to the scale invariance, we have that $4V_0(\mathbf{n})=\mathbf{n}\cdot \nabla_{\mathbf{N}}V_0(\mathbf{N})|_{\mathbf{N}=\mathbf{n}}=0$.

Finally, the $n^2_i$ in Eqs. (\ref{nsquared11})$-$(\ref{nsquared}) are a true local minimum if the Hessian matrix, $\textrm{P}|_{\mathbf{N}=\mathbf{n}}=\nabla_{\mathbf{N}}\nabla^T_{\mathbf{N}}V_0(\mathbf{N})|_{\mathbf{N}=\mathbf{n}}$, is positive semidefinite on the tangent space of the unit hypersphere at $\mathbf{N}=\mathbf{n}$. In particular, we have that $\textrm{P}$ is 
\begin{eqnarray}\label{Hessian}
\textrm{P}&=&\left(
\begin{array}{cccc}
\textrm{P}_{11} & \lambda_{\chi \rho} 
   n_\rho n_\chi & \lambda_{\chi \sigma} 
   n_\sigma n_\chi & \lambda_{\chi \varphi }
   n_\varphi n_\chi \\
 \lambda_{\chi \rho} n_\rho n_\chi &
  \textrm{P}_{22}  & \lambda_{\rho \sigma} 
   n_\rho n_\sigma & \lambda_{\rho \varphi} 
   n_\rho n_\varphi \\
 \lambda_{\chi \sigma} n_\sigma n_\chi &
   \lambda_{\rho \sigma} n_\rho n_\sigma &
    \textrm{P}_{33}  & \lambda_{\varphi \sigma} 
   n_\sigma n_\varphi  \\
 \lambda _{\chi \varphi} n_{\varphi} n_\chi &
   \lambda_{\rho \varphi}  n_{\rho} n_{\varphi} 
   & \lambda_{\varphi \sigma}  n_\sigma 
   n_\varphi &\textrm{P}_{44} 
\end{array}
\right),
\end{eqnarray}
where
\begin{eqnarray}
\textrm{P}_{11}&=&\frac{1}{2} \left(\lambda_{\chi \rho} n_\rho^2+\lambda_{\chi \sigma}  n_\sigma^2+\lambda_{\chi \varphi}  n_\varphi^2+6 \lambda_\chi n_\chi^2\right),  \quad \textrm{P}_{22}= \frac{1}{2}\left(6 \lambda_\rho  n_\rho^2+\lambda_{\rho \sigma}  n_\sigma^2+\lambda_{\rho \varphi} n\varphi^2+\lambda_{\chi \rho }n_\chi^2\right), \nonumber \\\
\textrm{P}_{33}&=&\frac{1}{2} \left(\lambda_{\rho \sigma}  n_\rho^2+6 \lambda_\sigma  n_\sigma ^2+\lambda_{\varphi \sigma}  n_\varphi^2+\lambda_{\chi \sigma} n_\chi^2\right), \quad \textrm{P}_{44}=\frac{1}{2} \left(\lambda_{\rho
   \varphi}  n_\rho^2+\lambda_{\varphi \sigma} 
   n_\sigma^2+6 \lambda_{\varphi } n_\varphi^2+\lambda_{\chi \varphi } n_\chi ^2\right). 
\end{eqnarray}
Therefore,  $\textrm{P}$ is positive semidefinite if and only if 
\begin{eqnarray}
&& \lambda_\chi \geq 0,\quad \lambda_\rho  \geq 0,\quad \lambda_\sigma \geq 0,\quad \lambda_\varphi \geq 0, \nonumber \\
&&  4 \lambda_\rho  \lambda_\chi  \geq \lambda_{\chi \rho}^2, \quad 4 \lambda_\sigma  \lambda_\chi \geq \lambda_{\chi \sigma}^2, \quad  4 \lambda_\varphi  \lambda_\chi \geq \lambda_{\chi \varphi}^2,  \nonumber \\
&& 4 \lambda_\rho  \lambda_\sigma \geq \lambda_{\rho \sigma}^2, \quad 4 \lambda_{\rho}  \lambda_{\varphi}\geq \lambda_{\rho \varphi}^2, \quad 4 \lambda_{\sigma}  \lambda_{\varphi}\geq \lambda_{\varphi \sigma}^2,   \nonumber \\
&&4 \lambda_\rho  \lambda_\sigma  \lambda_\chi +\lambda_{\rho
   \sigma}  \lambda_{\chi \rho}  \lambda_{\chi \sigma} \geq \lambda
   _\rho  \lambda_{\chi \sigma} ^2+\lambda _{\rho \sigma} ^2
   \lambda _\chi +\lambda _{\sigma}  \lambda_{ \chi \rho} ^2,   \nonumber \\
&& 4 \lambda_\rho  \lambda _\varphi  \lambda _\chi +\lambda _{\rho
   \varphi}  \lambda_{\chi \rho}  \lambda_{\chi \varphi} \geq \lambda
   _\rho  \lambda_{\chi \varphi}^2+\lambda_{\rho \varphi}^2
   \lambda_\chi +\lambda_{\varphi}  \lambda_{\chi \rho}^2,   \nonumber \\
&&  4 \lambda_{\sigma}  \lambda_{\varphi}  \lambda_\chi +\lambda
   _{\varphi \sigma}  \lambda_{\chi \sigma}  \lambda_{\chi \varphi}
   \geq \lambda_\sigma  \lambda_{\chi \varphi}^2+\lambda_{\varphi} 
   \lambda_{\chi \sigma}^2+\lambda_{\varphi \sigma}^2 \lambda_{
   \chi},   \nonumber \\
&& 4 \lambda_{\rho}  \lambda_{\sigma } \lambda_{\varphi} +\lambda
   _{\rho \sigma}  \lambda_{\rho \varphi}  \lambda_{\varphi \sigma}
   \geq \lambda_{\rho}  \lambda_{\varphi \sigma}^2+\lambda_{\rho
   \sigma}^2 \lambda_{\varphi} +\lambda_{\rho \varphi}^2
   \lambda_{\sigma},   \nonumber \\ 
&&\mathrm{det}\, {\bf{\Lambda}}_0\geq0,
\label{semipositive}
\end{eqnarray}
Notice that from Eq. \eqref{lambdasigma}, we have that $\mathrm{det}\, {\bf{\Lambda}}_0=0$ in such a way that the last condition of Eq. (\ref{semipositive}) is automatically satisfied. It is also important to compare the conditions coming from the vacuum stability, Eq. \eqref{cop_1}, to the ones coming from the positive semidefiniteness of the Hessian matrix $\mathrm{P}$, Eq. \eqref{semipositive}. For  $\lambda_{\chi \rho}^\prime>0$, the conditions in Eq. \eqref{cop_1} are automatically satisfied provided the conditions in Eq. \eqref{semipositive} are. This happens because the positive semidefinite matrices are a subset of the copositive matrices.  However, for $\lambda_{\chi\rho}^\prime< 0$, the matrix that governs the scalar potential behaviour in the limit of large fields is  ${\bf{\Lambda}}(\theta=1)$ instead of ${\bf{\Lambda}}_0$. Hence, both conditions, Eqs. \eqref{cop_1} and \eqref{semipositive}, must be simultaneously considered.

We now turn our attention to finding the scalar mass spectrum. The three CP-odd fields in Eq. (\ref{scalars}): $A_\rho$, $A_{\chi_3}$ and $A_\sigma$, remain massless and are absorbed by the three real neutral gauge bosons in the model: $Z,\, Z^\prime$ and $Z^{\prime\prime}$, via the Higgs mechanism. 

The four CP-even fields, in the basis ${\bf B}_S=(S_\rho, S_{\chi_3}, S_\sigma, S_\varphi)$, share the following squared mass matrix
{
\footnotesize
\be\label{MS}
M_S^2=\left(
\begin{array}{cccc}
 2 \lambda_\rho  v_\rho^2 & \lambda_{\rho \chi}  v_\rho v_\chi & \lambda_{\rho \sigma}  v_\rho v_\sigma & -\frac{v_\rho \left(2 \lambda_\rho  v_\rho^2+\lambda_{\rho \sigma} v_\sigma^2+\lambda_{\rho \chi}  v_\chi^2 \right)}{v_\varphi} \\
 \star & 2 \lambda_\chi  v_\chi^2 & \lambda_{\chi \sigma}  v_\sigma v_\chi & -\frac{v_\chi \left(\lambda_{\rho \chi}  v_\rho^2+\lambda_{ \chi \sigma}  v_\sigma^2+2 \lambda_\chi  v_\chi^2\right)}{v_\varphi} \\
 \star & \star & -\lambda_{\rho \sigma}  v_\rho^2-\lambda_{\sigma \varphi}  v_\varphi^2-\lambda_{\chi \sigma}  v_\chi^2 & \lambda_{\sigma \varphi}  v_\sigma v_\varphi \\
 \star & \star & \star & \frac{2 \lambda_\rho  v_\rho^4+v_\rho^2 \left(\lambda_{\rho \sigma}  v_\sigma^2+2 \lambda_{\rho \chi}  v_\chi^2\right)+v_\sigma^2 \left(\lambda_{\chi \sigma}  v_\chi^2-\lambda_{\sigma \varphi}  v_\varphi^2\right)+2 \lambda_\chi  v_\chi^4}{v_\varphi^2} \\
\end{array}
\right)
\ee
}
Note that the vacuum expectation values $v_{\rho},\,v_{\chi},\,v_{\sigma}$ and $v_{\varphi}$ are related with the $n_i$ given in Eqs. \eqref{nsquared11}$-$(\ref{nsquared}) through $v_{\rho}=\sqrt{2}n_\rho \langle \phi_r \rangle$, $v_{\chi}=\sqrt{2}n_\chi\langle \phi_r \rangle$, $v_{\sigma}=\sqrt{2}n_\sigma\langle \phi_r \rangle$ and $v_{\varphi}=\sqrt{2}n_\varphi\langle \phi_r \rangle$, where $\langle \phi_r \rangle$ is the breaking scale of scale invariance. As expected,
this matrix has a vanishing eigenvalue associated with the breaking of scale invariance. The corresponding massless state can be identified by performing the following unitary transformation ${B}_S \to U_{S1} {B}_{S1}$, where
\be \label{US1}
U_{S1}=
\left(
\begin{array}{cccc}
 \frac{v_\rho}{\sqrt{v_\rho^2 + v_\chi^2 + v_\sigma^2+ v_\varphi^2}} & \frac{v_\chi}{\sqrt{v_\rho^2 + v_\chi^2}} & \frac{v_\rho v_\sigma}{\sqrt{(v_\rho^2 + v_\chi^2) (v_\rho^2 + v_\chi^2 + v_\sigma^2)}} & \frac{v_\rho v_\varphi}{\sqrt{(v_\rho^2 + v_\chi^2 + v_\sigma^2) (v_\rho^2 + v_\chi^2 + v_\sigma^2+ v_\varphi^2)}} \\
 \frac{v_\chi}{\sqrt{v_\rho^2 + v_\chi^2 + v_\sigma^2+ v_\varphi^2}} & -\frac{v_\rho}{\sqrt{v_\rho^2 + v_\chi^2}} & \frac{v_\sigma v_\chi}{\sqrt{(v_\rho^2 + v_\chi^2) (v_\rho^2 + v_\chi^2 + v_\sigma^2)}} & \frac{v_\varphi v_\chi}{\sqrt{(v_\rho^2 + v_\chi^2 + v_\sigma^2) (v_\rho^2 + v_\chi^2 + v_\sigma^2+ v_\varphi^2)}} \\
 \frac{v_\sigma}{\sqrt{v_\rho^2 + v_\chi^2 + v_\sigma^2+ v_\varphi^2}} & 0 & -\frac{v_\rho^2 + v_\chi^2}{\sqrt{(v_\rho^2 + v_\chi^2) (v_\rho^2 + v_\chi^2 + v_\sigma^2)}} & \frac{v_\sigma v_\varphi}{\sqrt{(v_\rho^2 + v_\chi^2 + v_\sigma^2) (v_\rho^2 + v_\chi^2 + v_\sigma^2+ v_\varphi^2)}} \\
 \frac{v_\varphi}{\sqrt{v_\rho^2 + v_\chi^2 + v_\sigma^2+ v_\varphi^2}} & 0 & 0 & -\frac{v^2+v_\sigma^2+v_\chi^2}{\sqrt{(v_\rho^2 + v_\chi^2 + v_\sigma^2) (v_\rho^2 + v_\chi^2 + v_\sigma^2+ v_\varphi^2)}} \\
\end{array}
\right).
\ee
The massless field, the pseudo-Goldstone boson of the scale symmetry, gets its components from the first column of $U_{S1}$:
\be
S = \frac{1}{\sqrt{v_\rho^2 + v_\chi^2 + v_\sigma^2 + v_\varphi^2}}\left( v_\rho S_\rho + v_\chi S_{\chi_3} +v_\sigma S_\sigma +v_\varphi S_\varphi \right)\,,
\ee 
and assuming the limit $v_\rho \ll v_\chi \ll v_\sigma,v_\varphi$, we see that the dominant contributions to $S$ fall along the singlet field components $S_\sigma, S_\varphi$. Among the remaining three CP-even fields, the lightest one gets a mass proportional to the electroweak vev and is identified with the SM Higgs boson. It  is approximately given by the second column of $U_{S1}$:
\be
h\simeq \frac{1}{\sqrt{v_\rho^2+v_\chi^2}}\left( v_\chi S_\rho - v_\rho S_{\chi_3}\right)\,.
\ee 
The other two physical states, $H_1$ and $H_2$, whose main contributions fall along $S_{\chi_3}$ and $S_\sigma,S_\varphi$ get heavy masses proportional to $v_\chi$ and $v_\sigma\simeq v_\varphi$, respectively. 

The last neutral field, the complex scalar $\chi_1^0=\frac{S_{\chi_1}+iA_{\chi_1}}{\sqrt{2}}$, remains massless and is absorbed by the complex vector boson $Y^0$ (see Section \ref{sec:gauge}). Meanwhile, amongst the charged fields, we have two pseudo-Goldstones, $G_W^\pm$ and $G_{W^\prime}^\pm$, eaten by the gauge bosons $W^\pm$ and $W^{\prime\pm}$, in addition to a massive field $H^\pm$. These fields, in the mass basis, are given by
\bea
G_{W}^{\pm} &=& \rho_1^\pm \,,\quad \quad G_{W^\prime}^{\pm} = \frac{1}{v_\rho^2 + v_\chi^2} ( -v_\rho \rho_3^\pm + v_\chi \chi_2^\pm), \quad H^\pm = \frac{1}{v_\rho^2 + v_\chi^2} ( v_\chi \rho_3^\pm + v_\rho \chi_2^\pm),
\eea
where the squared mass of $H^\pm$ is
\be 
m_{H^\pm}^2 = \frac{\lambda^\prime_{\rho\chi}}{2}( v_\rho^2+v_\chi^2 ). \label{masahpm}
\ee
From Eq. (\ref{masahpm}), we notice that unless $\lambda^\prime_{\chi \rho}>0$, we would have a tachyonic field. Because $\lambda^\prime_{\chi \rho}>0$ we also have that the necessary and sufficient conditions for vacuum stability are those shown in Eq. (\ref{semipositive}).

\section{Gauge sector}\label{sec:gauge}

The gauge bosons become massive as a consequence of the symmetry breaking taking place in the scalar sector. Therefore, we can obtain their masses from the terms in the Lagrangian that contain the covariant derivatives of the scalars as they lead to interactions between gauge bosons and scalars. The $\text{SU}(3)_L\otimes \text{U}(1)_X \otimes \text{U}(1)_N$ covariant derivative, acting on a generic $\text{SU}(3)_L$ triplet $\eta$, is given by 
\be\label{cov_der}
D_\mu\, \eta = \left( \partial_\mu - i g_L W^a_\mu T_a - ig_X X_{\eta} X_\mu- ig_N N_{\eta} N_\mu\right) \eta,
\ee 
where $T_a=\lambda_a/2$ are the generators of $\text{SU}(3)_L$ (with $\lambda_a$ being the well-known Gell-Mann matrices) and $X_{\eta}$ ($N_{\eta}$) the $\eta$ charge under $\text{U}(1)_X$ ($\text{U}(1)_N$); whereas $W^a_\mu$, $X_\mu$ and $N_\mu$ are the $\text{SU}(3)_L$, $\text{U}(1)_X$, $\text{U}(1)_N$ gauge fields, respectively.

The neutral gauge bosons, in the diagonal of the covariant derivative defined in Eq. (\ref{cov_der}), mix among themselves once the scalars $\rho$, $\chi$ and $\sigma$ acquire vevs. Choosing ${\bf B}_{\mu}=(W_\mu^3, W^8_\mu, X_\mu, N_\mu)^T$ as our basis, the corresponding squared mass matrix is 
\be 
M^2=g_L^2\left(
\begin{array}{cccc}
 \frac{v_\rho^2}{4} & \star & \star & \star \\
 -\frac{v_\rho^2}{4 \sqrt{3}} & \frac{1}{12} \left(v_\rho^2+4 v_\chi^2\right) & \star & \star \\
 -\frac{t_X v_\rho^2}{3} & \frac{t_X \left(v_\rho^2+v_\chi^2\right)}{3 \sqrt{3}} & \frac{t_X^2}{9}  \left(4 v_\rho^2+v_\chi^2\right) & \star \\
 -\frac{t_N v_\rho^2}{3}  & \frac{t_N \left(v_\rho^2+4 v_\chi^2\right)}{3 \sqrt{3}} & \frac{4t_N t_X}{9}  \left(v_\rho^2+v_\chi^2\right) & \frac{4t_N^2}{9}  \left(v_\rho^2+9 v_\sigma^2+4 v_\chi^2\right) \\
\end{array}
\right)\,,
\ee 
where $t_{X(N)} \equiv g_{X(N)}/g_L$. As usual, we can diagonalise $M^2$ by means of a unitary change of basis ${\bf B_\mu} \to U{\bf B_\mu}^\prime$, where ${\bf B}_{\mu}^\prime=(A_\mu, Z_\mu, Z^\prime_\mu, Z^{\prime\prime}_\mu)$ is the mass basis, so that $U^T M^2 U= \mbox{diag}(m)$. We break the diagonalisation procedure in three steps $U=U_1U_2U_3$ with
\bea 
U&=&\left(
\begin{array}{cccc}
 \frac{\sqrt{3} t_X}{\sqrt{4 t_X^2+3}} & 
 \frac{t_X^2+3}{\sqrt{\left(t_X^2+3\right) \left(4 t_X^2+3\right)}}  & 0 & 0 \\ -\frac{t_X}{\sqrt{4 t_X^2+3}} & \frac{\sqrt{3} t_X^2}{\sqrt{\left(t_X^2+3\right) \left(4 t_X^2+3\right)}} & \frac{\sqrt{3}}{\sqrt{t_X^2+3}} & 0 \\
 \frac{\sqrt{3}}{\sqrt{4 t_X^2+3}} & -\frac{3 t_X}{\sqrt{\left(t_X^2+3\right) \left(4 t_X^2+3\right)}} & \frac{t_X}{\sqrt{t_X^2+3}} & 0 \\
 0 & 0 & 0 & 1 \\
\end{array}
\right)\,\,\left(
\begin{array}{cccc}
1 & 0 & 0 & 0 \\
0 & 1 & \epsilon_1 & \epsilon_2 \\
0 & -\epsilon_1 & 1 & 0 \\
0 & -\epsilon_2 & 0 & 1 \\
\end{array}
\right)\,\left(
\begin{array}{cccc}
1 & 0 & 0 & 0 \\
0 & 1 & 0 & 0 \\
0 & 0 & c_\theta & -s_\theta\\
0 & 0 & s_\theta & c_\theta \\
\end{array}
\right)\,,\nn\\
\mbox{where}&&\,\,\epsilon_1 \simeq-\frac{v^2_\rho\left[3 v_\sigma^2\left(4 t_X^2+3\right) +4 t_X^2 v^2_\chi\right]\sqrt{4 t_X^2+3} }{v^2_\rho v_\sigma^2 \left(4 t_X^2+3\right)^2+4 v^2_\chi \left(t_X^4 v_\rho^2+\left(t_X^2+3\right)^2 v_\sigma^2\right)}\,,\\
&&\,\,\epsilon_2 \simeq \frac{ v^2_\rho v^2_\chi t_X^2 \sqrt{(t_X^2+3)(4 t_X^2+3)} }{v^2_\rho v_\sigma^2 t_N\left(4 t_X^2+3\right)^2 +4 t_N v^2_\chi \left[t_X^4 v_\rho^2+\left(t_X^2+3\right)^2 v_\sigma^2\right]}\,,\nonumber\\
\mbox{and,}&&\\
&&\,\,\tan{\left(2\theta\right)} \simeq 
\frac{8v_\chi^2 t_N   \sqrt{t_X^2+3}}{\left(t_X^2+3\right) v_\chi^2-4 t_N^2 \left(9 v_\sigma^2+4 v_\chi^2\right)}.\nn
\eea
In the first step, $U_1$, we identify the photon, the only massless state, as
\be 
A_\mu = \frac{1}{\sqrt{3+4 t_X^2}}\left(\sqrt{3} t_X W^3_\mu -t_X  W^8_\mu + \sqrt{3}  X_\mu \right).
\ee 
The resulting matrix, $M_{1}^2=U_1^T M^2 U_1$, omitted here for conciseness, contains a non-diagonal $3\times 3$ block which mixes the transformed fields, except for the photon. We notice that a $2\times2$ sub-block of $M_1^2$ is governed by the large vevs $v_\chi$ and $v_\sigma$, whereas the other non-vanishing entries are governed by $v_\rho (\ll v_\chi,v_\sigma)$; therefore, in the second step, we employ a seesaw approximation via $\epsilon_{1,2}\ll 1$ to separate the light state from the other two and obtain $M_{2}^2=U_2^T M_1^2U_2$. The lightest among the massive neutral bosons is identified with the SM $Z_\mu$ boson and its mass is given by
\be 
m_Z \simeq \frac{g_Lv_\rho}{2} \sqrt{\frac{4 t_X^2+3}{t_X^2+3}}.
\ee 
As a third and last step, we perform another unitary transformation, $U_3$, to diagonalise the $2\times 2$ block in $M_{2}^2$ that mixes the two heavy neutral bosons. The two heavy states, $Z_\mu^\prime$ and $Z_\mu^{\prime\prime}$, obtain the following squared masses 
\bea \label{zzmass}
m_{Z^\prime,Z^{\prime\prime}}^2 &\simeq& \frac{1}{18} g_L^2 \left(36 t_N^2 v_\sigma^2 + v_\chi^2 \left(16 t_N^2+t_X^2+3\right)\mp6 t_N v_\sigma \sqrt{36 t_N^2 v_\sigma^2+2 v_\chi^2 \left(16 t_N^2-t_X^2-3\right)}\right)\,.
\eea

Finally, the remaining fields become three massive complex vector bosons which remain unmixed: $Y^{0(\dagger)}_\mu$, $W^{\pm}_\mu$, $W^{\prime\pm}_\mu$, the first of which is electrically neutral, while the other two are singly charged. These gauge bosons and their masses are defined by
\be
Y^{0(\dagger)}_\mu = \frac{W_\mu^4 \mp i W_\mu^5}{\sqrt{2}} \, ,\quad \quad W^\pm_\mu = \frac{W_\mu^1 \mp i W_\mu^2}{\sqrt{2}}\,,\quad \quad 
W^{\prime\pm}_\mu = \frac{W_\mu^6 \mp i W_\mu^7}{\sqrt{2}}\,,
\ee
and
\be 
m_{Y^0} = \frac{g_L v_\chi}{2}\,,\quad \quad m_{W^\pm} = \frac{g_L v_\rho}{2}\, ,\quad \quad m_{W^{\prime\, \pm}} = \frac{g_L \sqrt{v_\rho^2+v_\chi^2} }{2}\,,
\ee 
respectively.
Notice that $W^\pm_\mu$ is the charged SM vector boson, while $Y^{0(\dagger)}_\mu$ and $W^{\prime\pm}_\mu$ represent heavier gauge bosons.

\section{Fermion masses} \label{fermionmassessection}

The minimality of the 3-3-1 model discussed in Sec. \ref{minimal} is such that some of its fermions remain massless. In what follows, we show how with the introduction of the fields in Eqs. (\ref{newq}), (\ref{newl}) and (\ref{news}) all fermions become massive in our extended construction. In particular, we discuss the emergence of mass hierarchies among fermions of the same type via seesaw mechanisms.

\subsection{Quark sector}\label{subsecQuarks}

We start by writing down all the allowed renormalisable interactions between quarks and scalars, {\it i.e.:}
\begin{eqnarray}\label{Yukq}
-\mathcal{L}^{Y}_{q}&=&Y^{d}_{1 i}\overline{Q_{1L}}d_{ i R}\rho+Y^{u}_{a i}\overline{Q_{a L}}u_{i R}\rho^{*}+Y^{U}_{1}\overline{Q_{1L}}U_{R}\chi+Y^{D}_{a b}\overline{Q_{a L}} D_{b R}\chi^{\ast}+y^{{\cal{Q}}_1}\overline{Q_{1L}}{\cal{Q}}_R\rho^{*} \nonumber\\
&&+y^{u}_i\overline{{\cal{Q}}_L}u_{i R}\chi^{*} + y_{a b}^{{\cal{Q}}_2} \overline{Q_{aL}}{\cal{Q}}_{b R}\rho+y^d_{ai}\overline{{\cal{Q}}_{a L}}d_{i R}\chi+ h^{{\cal{Q}}_1}\overline{{\cal Q}_{L}} {\cal{Q}}_{R}\varphi+h^{{\cal{Q}}_2}_{ab}\overline{{\cal Q}_{aL}} {\cal{Q}}_{bR}\varphi+\textrm{H.c.\,,}
\end{eqnarray}
where the $Y$ yukawas are associated with interactions already present in the minimal model, $y$ yukawas represent interactions that mix fields in the minimal model with the new fields, and $h$ yukawas govern interactions involving only the new fields.
Once the scalars acquire non-vanishing vevs, quark mass terms are generated. Considering only the up-type quarks, two independent mass matrices are generated, 
\begin{equation}\label{mus}
M_{1}^u\equiv \left(
\begin{array}{c}
m_{u[3\times 4]} \\
m_{U[1\times 4]}
\end{array}
\right)
=\frac{1}{\sqrt{2}}
\left(
\begin{array}{cc}
0_{[1\times 3]} &-y^{{\cal{Q}}_1}v_{\rho}\\
Y^{u}_{[2 \times 3]}v_{\rho}  & 0_{[2\times 1]}\\
y^{u}_{[1\times 3]}v_\chi & h^{{\cal{Q}}_1}v_{\varphi}
\end{array}
\right) 
\text{,\,\,\,\,}M^u_{2}=\frac{1}{\sqrt{2}}
\left(
\begin{array}{cc}
Y^{U}_{1}v_\chi &y^{{\cal{Q}}_1}v_{\rho}\\
0 & h^{{\cal{Q}}_1}v_{\varphi} 
\end{array}
\right),
\end{equation}
where the bases $\mathbf{u_1}_{L,R}\equiv\left(u_{1},u_{2},u_{3},{\cal{U}}_{2}\right)_{L,R}$ and $\mathbf{u_2}_{L,R}\equiv\left(U,{\cal{U}}_{1}\right)_{L,R}$, respectively, have been assumed. The quarks in both bases carry the same electric charges of $2/3$. Nevertheless, the fields in $\mathbf{u_1}_{L,R}$ do not mix with those in $\mathbf{u_2}_{L,R}$ because they have different $B-L$ charges. The quarks associated with $M^{u}_{2}$ become very heavy with masses proportional to $v_\chi$ and $v_\varphi$. Regarding $M_{1}^u$ -- which, for later convenience, can be divided into $m_u$ and $m_U$ -- we first point out that its determinant does not vanish automatically, implying that the associated fields are all massive. Moreover, taking into account the vev hierarchy $v_\varphi \gg v_\chi \gg v_\rho$, we notice that the mass matrix presents a seesaw-like texture. In order to obtain the mass eigenvalues -- or at least their corresponding effective mass scales -- we adopt the approach presented in Ref.~\cite{Grimus:2000vj} applied to the hermitian matrix $M^u_1 M_1^{u\dagger}$. The first step converts $M^u_1 M_1^{u\dagger}$ into the block-diagonal matrix, via a unitary transformation $U_u$, as shown below
\be\label{Uu}
U_u^\dagger M^u_1 M_1^{u\dagger} U_u =
\begin{pmatrix}
D^{\text{light}}_{u[3\times3]} & 0_{[3\times1]}\\
0_{[1\times3]} & D^{\text{heavy}}_{u[1\times 1]} 
\end{pmatrix}~,\quad \quad \mbox{with}\quad \quad U_{u} = 
\begin{pmatrix}
\left(1-F_u F^\dagger_u\right)^{1/2}_{[3\times 3]} & F_{u[3\times 1]} \\
- \left(F_{u}^\dagger\right)_{[1\times 3]} & \left(1-F_u^\dagger F_u\right)^{1/2}_{[1\times 1]} 
\end{pmatrix}~,
\ee 
where the $3\times 1$ matrix $F_u$ can be written as $F_u \simeq m_u m_U^\dagger(m_U m_U^\dagger)^{-1}$, with $m_u$ and $m_U$ defined in Eq. (\ref{mus}). The non-vanishing matrices can be written as
\be 
D^{\textrm{light}}_{u[3\times3]}  
 \simeq \frac{v_\rho^2}{2}
\left(
\begin{array}{cc}
 \frac{ |y^{\mathcal{Q}_1}|^2 \kappa^2}{|h^{{\cal{Q}}_1}|^2} (y^u y^{u\dagger})_{[1\times1]}  &\frac{ y^{\mathcal{Q}_1}\kappa}{h^{{\cal{Q}}_1}} (y^u Y^{u\dagger})_{[1\times2]}\\
 \frac{ y^{\mathcal{Q}_1*} \kappa}{ h^{{\cal{Q}}_1*}}  (Y^u y^{u\dagger})_{[2\times1]} &  (Y^u Y^{u\dagger})_{[2\times2]}
 \\
\end{array}
\right)~, \quad \quad
D^{\textrm{heavy}}_{u[1\times1]}\simeq \frac{v_\varphi^2}{2}|h^{{\cal{Q}}_1}|^2\,, \label{upqmasses}
\ee
where $\kappa =v_\chi/v_\varphi\ll1$. We can repeat the block-diagonalisation procedure now with $D^{\text{light}}_{u[3\times 3]}$ to split it into the following two blocks governed by different mass scales
\be
m_{u_1}^2 
 \simeq \frac{v_\rho^2 \kappa^2}{2} \frac{|y^{{\cal{Q}}_1}|^2}{|h^{\mathcal{Q}_1}|^2} [y^u(1-Y^\dagger(Y Y^\dagger)^{-1}Y)y^{u\dagger}]_{[1\times 1]}~, \quad \quad
m^2_{u_{23}[2\times2]}\simeq \frac{v_\rho^2}{2}(Y Y^\dagger)_{[2\times 2]}\,.
\ee
From the first entry, we obtain the mass of the lightest up-type quark, $m_{u_1}$, which is suppressed by a factor $\kappa = v_\chi/v_\varphi$ with respect to the electro-weak scale, $v_\rho$. Whereas from the second term, $m^2_{u_{23}[2\times2]}$, we obtain the masses of the second and third up-type quarks, both proportional to $v_\rho$ with no suppression scale. Therefore, the model predicts a mass hierarchy between the up-type quark generations, with the first generation being lighter than the others.

Moving on to the down-type quarks, {\it i.e.} quarks with an electric charge of $-1/3$, we can also define two independent mass matrices, in the bases $\mathbf{d_1}_{L,R}\equiv\left(d_{1},d_{2L},d_{3},{\cal{D}'}_{2},{\cal{D}'}_{3}\right)_{L,R}$ and $\mathbf{d_2}_{L,R}\equiv\left(D_{2},D_{3},{\cal{D}}_{2},{\cal{D}}_{3}\right)_{L,R}$,  as
\begin{equation}\label{mds}
M^d_1\equiv
\left(
\begin{array}{c}
m_{d[3\times 5]} \\
m_{D[2\times 5]}
\end{array}
\right)
=
\frac{1}{\sqrt{2}}
\left(
\begin{array}{cc}
Y^{d}_{[1\times 3]}v_\rho & 0_{[1\times2]}\\
0_{[2\times 3]} & -y^{{\cal{Q}}_2}_{[2\times2]}v_\rho \\
y^{d}_{[2\times 3]}v_{\chi} & h^{{\cal{Q}}_2}_{[2\times 2]}v_{\varphi} 
\end{array}
\right),\quad \quad
M^d_2=\frac{1}{\sqrt{2}}
\left(
\begin{array}{cc}
Y^{D}_{[2\times 2]}v_\chi & y^{{\cal{Q}}_2}_{[2\times2]} v_\rho\\
0_{[2\times2]} & h^{{\cal{Q}}_2}_{[2\times2]} v_{\varphi}  \\
\end{array}
\right), 
\end{equation}
where the fields in $\mathbf{d_1}_{L,R}$ and $\mathbf{d_2}_{L,R}$ carry different $B-L$ charges and, as such, do not mix.
The matrix $M^d_2$ gives rise to four very large masses, two of which are proportional to $v_\chi$, while the remaining two are governed by $v_\varphi$.
Following the same procedure adopted for the up-type quarks in Eq. (\ref{Uu}), we can block-diagonalise $M^d_1$ via a $5\times 5$ unitary transformation $U_d$, written as a function of $F_{d[3\times 2]}\simeq m_d m_D^\dagger (m_D m_D^\dagger)^{-1}$. This will lead to a $3\times 3$ matrix for the light quarks and a $2\times2$ matrix for the heavy quarks:
\bea
D^{\textrm{light}}_{d[3\times3]}  
 \simeq \frac{v_\rho^2}{2}
\left(
\begin{array}{cc}
 (Y^{d} Y^{d\dagger})_{[1\times1]}  & \kappa(Y^d A^\dagger)_{[1\times2]} \\ \kappa (A Y^{d\dagger} )_{[2\times1]}  & \kappa^2 (AA^\dagger)_{[2\times2]} 
 \\
\end{array}
\right)~, \quad \quad
D^{\textrm{heavy}}_{d[2\times2]}\simeq \frac{v_\varphi^2}{2}( h^{{\cal{Q}}_2} h^{{{\cal{Q}}_2}\dagger})_{[2\times2]}\,, \label{downqmasses}
\eea
where $A_{[2\times3]}=y^{\mathcal{Q}_2} (h^{{\cal{Q}}_2})^{-1}y^{d}$.
Repeating the block-diagonalisation procedure for $D^{\text{light}}_{d[3\times3]}$, we obtain
\be
m^2_{d_1}
 \simeq\frac{v_\rho^2}{2}(Y^d Y^{d\dagger})_{[1\times 1]}~, \quad \quad
m^2_{d_{23}[2\times2]} \simeq  \frac{v_\rho^2 \kappa^2}{2}  [A(1-Y^{d\dagger}(Y^d Y^{d\dagger})^{-1}Y^d)A^\dagger]_{[2\times 2]}\,.
\ee
Therefore, we find that there is also a mass hierarchy among the different generations of down-type quarks. In this case, however, while the first generation gets a mass proportional to $v_\rho$, the second and third generation masses are proportional to $v_\rho \kappa$.

Therefore, we found above that the first generation of up-type and the second and third generations of down-type quarks -- {\it i.e.} the up, the strange and bottom quarks -- get masses suppressed by $\kappa$ with respect to the electro-weak scale $v_\rho \simeq {\cal{O}}(10^2)$ GeV. Because the heaviest among these fields, the bottom, has a GeV-scale mass, and also assuming that Yukawa couplings are at most of order unity, we have that $v_\rho \kappa \geq {\cal{O}}(1)$ GeV.  Thus, we adopt as a benchmark: $\kappa \simeq 10^{-2}$, which, for instance, can be satisfied for $v_\chi \simeq 10$ TeV ($\text{SU}(3)_L$-breaking scale) and $v_\varphi \simeq 10^3$ TeV.

In addition to the up- and down-type quarks, the quarks carrying non-SM electric charges, $q^{(5/3)}$ and $q^{(-4/3)}_a$, also become massive. These fields get the following mass terms, respectively,
\be\label{exqmass}
m^{(5/3)}= \frac{h^{{\cal{Q}}_1} v_\varphi}{\sqrt{2}}\quad\quad\mbox{and}\quad\quad
m^{(-4/3)}_{[2\times2]}= \frac{h^{{\cal{Q}}_2}_{[2\times2]} v_\varphi}{\sqrt{2}}.
\ee

\subsection{Reproducing the CKM matrix: a benchmark}

In this section, we show that it is indeed possible to obtain the masses and mixing for the standard quarks with the assumed energy scale for the vevs and a reasonable set of values for the Yukawa couplings. In order to give a numerical example with the vevs adopted previously -- {\it i.e.} $v_\rho = 246$ GeV, $v_\chi = 10$  TeV and $v_\sigma = 10^3$  TeV ($\kappa = v_\chi/v_\sigma= 10^{-2}$) -- we proceed to find the matrices $D^{\textrm{light}}_{u[3\times3]}$ and $D^{\textrm{light}}_{d[3\times3]}$, given in Eqs.~(\ref{upqmasses}) and (\ref{downqmasses}), respectively. These matrices are diagonalised by the unitary matrices $V^u_L$ and $V^d_L$, which satisfy the relation $V_{\textrm{CKM}} = V^u_L V^{d\dagger}_L$, where $V_{\textrm{CKM}}$ is the CKM matrix~\cite{ParticleDataGroup:2022pth}. 

As a first step, we generate a unitary mixing matrix for the standard up-type quarks, $V^u_L$, taken as  real, and then obtain the corresponding one for the standard down-type quarks through the relation $V^d_L=V_{\textrm{CKM}}^\dagger V^u_L$. Given that $V_L^{u\dagger} D^{\textrm{light}}_{u[3\times3]} V_L^u=\textrm{diag}(m_u^2,\,m_c^2,\,m_t^2)$, with $\textrm{diag}(m_u^2,\,m_c^2,\,m_t^2)=(4.6656\times10^{-6},\,1.6129,\,29821.8)$ GeV$^2$ for the central values for the up-type quark masses~\cite{ParticleDataGroup:2022pth}, we then find that $D^{\textrm{light}}_{u[3\times3]}=V_L^{u}\textrm{diag}(m_u^2,\,m_c^2,\,m_t^2)V_L^{u\dagger}$. Next, we check whether the newly-found matrix, $D^{\textrm{light}}_{u[3\times3]}$, can reproduce the expected hierarchy shown in Eq. (\ref{upqmasses}) with Yukawa coupling constants of order $ \mathcal{O}(0.1-1)$. As a benchmark, the unitary mixing matrix for the up-type quarks is 
\be
V^{u}_{L}=\left(
\begin{array}{ccc}
 0.8983994866 & 0.2975716999 & 0.3230006902 \\ -0.4391661419 & 0.6143740547 & 0.6554979944 \\ -0.0033855912 & -0.7307500286 &
0.6826367507 \\
\end{array}
\right).
\ee
This leads to the squared standard quark mass matrix 
\be
D^{\textrm{light}}_{u[3\times3]} =\left(
\begin{array}{ccc}
 0.6529034717 & 73.3446662997 & -69.3864182406\\
 73.3446662997 & 15925.3381887305 & -14875.5804906241\\
 -69.3864182406 & -14875.5804906241 & 13897.4579124635 \\
\end{array}
\right), 
\ee
which can then be obtained with the choices $y^{\mathcal{Q}_1}=0.1$, $h^{{\cal{Q}}_1}=0.1$, and the matrices 
\bea
y^{u}_{[1\times3]}&=&\left(
\begin{array}{ccc}
  0.1711600388 & -0.3782329121 & -0.2083816169 \\
\end{array}
\right),
\\
Y^{u}_{[2\times3]}&=&\left(
\begin{array}{ccc}
   0.1009743814 & -0.2180448060 & -0.6845282361 \\
  -0.0978206118 & 0.2113830498 & 0.6364329997 \\
\end{array}
\right).\nn
\eea
In addition, this benchmark implies that the heavy up-type quark squared mass in Eq. (\ref{upqmasses}) is $D^{\textrm{heavy}}_{u[1\times 1]} \simeq (71\,\, \mbox{TeV})^2$. 

Concerning the down-type quark sector, whose relevant matrices are shown in Eq. (\ref{downqmasses}), the squared mass matrix follows from $D^{\textrm{light}}_{d[3\times3]}=V_L^{d}\textrm{diag}(m_d^2,\,m_s^2,\,m_b^2)V_L^{d\dagger}$ where   $\textrm{diag}(m_d^2,\,m_s^2,\,m_b^2)=(2.18089\times10^{-5},\, 8.72356\times10^{-3},\,17.4724)$ GeV$^2$~\cite{ParticleDataGroup:2022pth}. In this case, the mixing matrix takes the form
{\footnotesize
\be
V^{d}_{L}=\left(
\begin{array}{ccc}
 0.9740839028 - 0.0000711801 i & 0.1459988055 - 0.0023054642 i &  0.1727257581 + 0.0023219427 i \\
 -0.2252470927 - 0.0000164370 i & 0.6950686946 - 0.0005323803 i & 
 0.6827464286 + 0.0005361856 i \\
 -0.0203761042 + 0.0030177189 i & -0.7039582806 + 0.0009995417 i & 
 0.7099410332 + 0.0010849575 i \\
\end{array}
\right),
\ee
}
and the squared mass matrix is
{\footnotesize
\be
D^{\textrm{light}}_{d[3\times3]} =\left(
\begin{array}{ccc}
0.0078766978 & 0.2493128589 + 0.0367627023 i & -0.2540335021 - 
  0.037820085 i \\
  0.2493128589 - 0.0367627023 i & 8.6628071391 & -8.7280022782 - 
  0.0257370553 i \\ -0.2540335021 + 0.037820085 i & -8.7280022782 + 
  0.0257370553 i & 8.8104615321\\ 
\end{array}
\right).
\ee
}
The latter can be reproduced by the following set of Yukawa constants: $y^{\mathcal{Q}_2}_{[2 \times 2]},h^{\mathcal{Q}_2}_{[2 \times 2]} =0.1\times 1_{[2 \times 2]}$ and
\be
Y^{d}_{[1\times3]}=\left(
\begin{array}{ccc}
   -0.2778517 & 0.2127718 + 0.2792035 i & 
  0.0745743 + 0.2330849 i \\
\end{array}
\right)\times 10^{-3},
\ee
{\footnotesize
\be
y^{d}_{[2\times3]}=\left(
\begin{array}{ccc}
  -0.8414096321 + 0.1307460931 i & 
  1.0644656013 + 0.7534651925 i & -0.0103744700 + 
   0.6610685237 i  \\
   0.8700762291 - 0.1444126405 i & -1.0350372937 - 0.7657291301 i 
   & -0.0436222750 - 
   0.6887273418 i \\
\end{array}
\right).
\ee}
Notice that, in contrast to the up-type quark case, here some Yukawa couplings need to be ``small'' -- that is, few orders of magnitude below 1 -- namely the elements of $Y^{d}_{[1\times3]}\sim\mathcal{O}(10^{-4})$. This happens because the mass hierarchy for the down-type quarks, given in Eq. (\ref{downqmasses}), is the opposite of that found for the up-type quarks. Finally, for the present benchmark, the heavy down-type quark masses, given by $D^{\textrm{heavy}}_{d[2\times 2]}$ in Eq. (\ref{downqmasses}), are the same and approximately equal to $71 \,\,\mbox{TeV}$. 

Lastly, it is worth mentioning that, in the present case, the mixing angles among standard and heavy quarks, governed by the matrices $F_u$ and $F_d$, defined below Eqs. (\ref{Uu}) and (\ref{mds}), for the up- and down-type quarks, respectively, are at most of order $v_\rho/v_\varphi\sim10^{-4}$; thus being very suppressed and well below the current upper limits \cite{ParticleDataGroup:2022pth}.

\subsection{Lepton sector \label{leptonsector}}

When it comes to the lepton fields, we can write down the following renormalisable Yukawa terms
\begin{eqnarray}\label{Yukl}
-\mathcal{L}^{Y}_{l}&=&Y^{e}_{i j}\overline{F_{i L}}e_{j R}\rho+Y^{\nu}_{i j}\overline{F_{i L}}\left(F_{j L}\right)^{C}\rho^{*}+ y^{\nu}_{ij}\overline{F_{i L}}\chi S_{j R} + \frac{1}{2} h^{S_R}_{ij} \sigma^*\overline{(S_{i R})^{C}} S_{j R}\\
&& + \frac{1}{2} h^{S_L}_{ij}\sigma \overline{S_{i L}} (S_{j L})^{C} + h^{S}_{ij} \overline{S_{i L}} S_{j R}\varphi
\,+\textrm{H.c.\,,}\nonumber
\end{eqnarray}
where, as already mentioned, the $\text{SU}(3)_L$ indices of the three (anti-)triplets are anti-symmetrically contracted to $\epsilon_{\alpha\beta\gamma}$ and, as a result, $Y^\nu_{ij}$ is also anti-symmetric.

The charged leptons, $e_i$, get their masses solely from the first term:
\be
M^e_{[3\times 3]} = \frac{v_\rho}{\sqrt{2}}Y^e_{[3\times 3]}\,. \label{emass}
\ee 
Notice that, in contrast to the quark sector, there is no seesaw mechanism for the charged leptons, and all generations get masses proportional to the same (electroweak) vev, $v_\rho$.

Finally, we look into the neutrino sector. As a result of introducing the extra fields in Eq. (\ref{newl}), the neutral lepton sector accommodates many degrees of freedom, which can be organised into the basis ${\bf {\tilde{N} } }_{L}=(\nu_{i \,L},(N_{iR})^{C},(S_{iR})^{C}, S_{iL})$ and are related to the mass matrix below:
\begin{eqnarray}\label{MN}
M^{\tilde{N}}=\frac{1}{\sqrt{2}}\left(
\begin{array}{cccc}
0 & 2Y^{\nu\,\textrm{T}}v_{\rho} & 0 & 0\\
2Y^{\nu}v_{\rho} & 0 & \text{y}^{\nu}v_{\chi} & 0\\
0 & \text{y}^{\nu\, \textrm{T}}v_{\chi} & h^{S_R}v_{\sigma} & h^{S\,\text{T}} v_\varphi\\
0 & 0 & h^{S}v_{\varphi} & h^{S_L}v_\sigma
\end{array}
\right),
\end{eqnarray}
where, for the sake of convenience, we are omitting the indices and dimensions of the Yukawa matrices, which are all $3\times3$.

The neutrino mass matrix above displays a double seesaw texture and hierarchy since $v_\sigma \simeq v_\varphi\gg v_\chi\gg v_\rho$.
Therefore, we divide the block-diagonalisation into two seesaw steps. First, we neglect the first block-row and block-column of the matrix in  Eq. (\ref{MN}), whose non-vanishing entries are proportional to $v_\rho$, and perform the block-diagonalisation of the remaining matrix.
As a result, it is easy to see that six  of the neutral leptons -- whose mass matrix is approximately given by the lower-right $6\times6$ sub-matrix of $M^{\tilde{N}}$ -- become very heavy with masses of the form $m_{S_i}=v_\varphi\tilde{h}^S_i/\sqrt{2}$, with $i=1,...,6$, where $\tilde{h}^S_i$ represent effective couplings. On the other hand, three neutral leptons will have seesaw masses suppressed by $v_\varphi$ with respect to the intermediate scale $v_\chi$, given by 
\be
m_{N} \simeq -\frac{v_\chi^2}{\sqrt{2}v_\sigma}\,y^{\nu}\left(h^{S_R}-h^{S\, \text{T}} (h^{S_L})^{-1} h^S \frac{v_\sigma}{v_\varphi}    \right)^{-1} y^{\nu\, \text{T}} \,.
\ee 
Supposing that the $h$ couplings are of order $10^{-1}$ and $y^\nu \simeq 1$, then $m_N = {\cal{O}}(1)$ TeV, while $m_S = {\cal{O}}(10^2)$ TeV.

Then the resulting $6\times 6$ upper-block of the matrix in Eq. (\ref{MN}) becomes
\be
M^{\tilde{N}}_{\text{upper}}\simeq\frac{1}{\sqrt{2}}\left( \begin{array}{cc}
0 & 2Y^{\nu\,\textrm{T}}v_{\rho}\\
2Y^{\nu}v_{\rho} & \sqrt{2}m_{N}
\end{array}\right). \label{neutrinosmasses} 
\ee 
Clearly, this matrix describes a type-I seesaw mass matrix for $m_N\gg Y^\nu v_\rho$ and can be diagonalised as such to find the masses of the neutrinos below
\be 
m_{\nu} \simeq - 2 v_\rho^2 \,Y^{\nu \,\text{T}} m_N^{-1} Y^\nu \,.
\ee 
Notice that the determinant of $m_\nu$ vanishes due to the anti-symmetric nature of $Y^\nu_{ij}$, implying that one neutrino is left massless. Thus, considering the benchmark choices so far adopted, we need $Y^\nu \leq 10^{-5}$ in order for the mass of the heaviest neutrino to be around or below the eV scale. Despite being smaller than the other couplings in the neutral lepton sector, $Y^\nu$ does not need to be smaller than the coupling required to reproduce the electron mass either in our model or in the SM.

\section{One-loop effective potential}\label{sec:disc}

In order to study the consistency of the symmetry breaking mechanism, we consider the first-order quantum corrections to the scalar potential in the 
$\phi_r \mathbf{n}$ direction, {\it i.e}. we calculate $V_{\mathrm{1-loop}}(\phi_r \mathbf{n})$, where $\mathbf{n}$ is the flat direction given by Eqs. \eqref{nchi}-\eqref{nsquared}. In the $\overline{\textrm{MS}}$ renormalisation scheme, the $V_{\mathrm{1-loop}}(\phi_r \mathbf{n})$ is 
\begin{eqnarray}
V_{\mathrm{1-loop}}(\phi_r\mathbf{n})=A \phi_r^{4}+B\,\phi_r^{4}\ln\left(\frac{\phi_r^{2}}{\mu_0^{2}}\right),
\label{vloop}
\end{eqnarray}
where $\mu_0$ is the same renormalisation scale that appears in Eq. \eqref{lambdasigma}, and the coefficients $A$ and $B$ are 
\begin{eqnarray}\label{CoeffA}
A=\frac{1}{64\pi^{2}\langle\phi_r\rangle^{4}}&& \left[ \sum_{\mathcal{S}}n_{\mathcal{S}}\,m_{\mathcal{S}}^{4}\left(\ln\frac{m_{\mathcal{S}}^{2}}{\langle\phi_r\rangle^{2}}-\frac{3}{2}\right)+ 3\sum_{\mathcal{V}}n_{\mathcal{V}}\,m_{\mathcal{V}}^4\left(\ln \frac{m_{\mathcal{V}}^{2}}{\langle\phi_r\rangle^2}-\frac{5}{6}\right)\right. \nonumber\\
&&\left.-4\sum_{ \mathcal{F}}\,n_{\mathcal{C}} n_\mathcal{M} \mathrm{Tr}\left[M_{\mathcal{F}}^{4}\left(\ln\frac{M_{\mathcal{F}}^{2}}{\langle\phi_r\rangle^2}-1\right)\right]\right],
\end{eqnarray}
and
\begin{eqnarray}
B&=& \frac{1}{64\pi^{2}\langle\phi_r\rangle^{4}}\left[\sum_{\mathcal{S}}n_{\mathcal{S}}\,m_{\mathcal{S}}^{4}+3\sum_{\mathcal{V}}n_{\mathcal{V}}\,m_{\mathcal{V}}^4-4\sum_{ \mathcal{F}}\,n_{\mathcal{C}} n_\mathcal{M} \mathrm{Tr}\left[M_{\mathcal{F}}^{4}\right]\right],
\label{CoeffB}
\end{eqnarray}
where $m_\mathcal{S},m_\mathcal{V}$ are the tree-level masses of the scalars $\mathcal{S}= h, H^{\pm}, H_1, H_2$ and vector bosons $\mathcal{V}=Z, Z', Z'', Y^{0(\dagger)}, W^{\pm}, W'^\pm$, respectively. $M_{\mathcal{F}}$ represents the mass matrices of the fermions, quarks and leptons, as given in Eqs. \eqref{mus}, \eqref{mds}, \eqref{exqmass}, \eqref{emass} and \eqref{MN}. We also have that $n_{\mathcal{S,\,V}}=2$ for $\mathcal{S}=H^{\pm}$ and $\mathcal{V}=W^{\pm}, W^{\prime\pm}, Y^{0(\dagger)}$ and equals $1$ otherwise. $n_{\mathcal{C}}=3$ for quarks and equals $1$ otherwise. Finally, $n_\mathcal{M} =1/2$ for Majorana fermions and $1$ otherwise.

The definite value of $\phi_r$ on the $\mathbf{n}$ direction, {\it{i.e.}} $\langle\phi_r\rangle$, that comes from the solution of $0=\left[ \frac{\partial V_{\mathrm{1-loop}}(\phi_r\mathbf{n})}{\partial\phi_r}\right]_{\langle\phi_r\rangle}$, is related to the renormalisation scale $\mu_0$ through $\langle\phi_r\rangle=\mu_0\exp{\left[-\frac{1}{4}-\frac{A}{2B}\right]}$. Using this relation to eliminate the explicit dependence on $\mu_0$ of $V_{\mathrm{1-loop}}(\phi_r\mathbf{n})$, we obtain
\be
V_{\mathrm{1-loop}}(\phi_r\mathbf{n})=B\,\phi_r^{4}\left[\ln\left(\frac{\phi_r^{2}}{\langle \phi_r\rangle^{2}}\right)-\frac{1}{2}\right],
\label{vloop2}
\ee
which is valid for $B\neq0$. From Eq. \eqref{vloop2} it is clear that $\langle\phi_r\rangle\mathbf{n}$ is not a minimum unless $B>0$. This condition on $B$ also brings a constraint on the masses of the particles in the model because from Eq. \eqref{CoeffB} we can see that fermion masses can not dominate over the boson masses.  Furthermore, as a consequence of the scale-invariance breaking, the scalon $S$ gets a squared mass equal to $8B\langle\phi_r\rangle^2$ which is positive if $B>0$.

In general, the calculation of coefficient $B$ in this model is not trivial because it involves the diagonalization of several mass matrices. However, we can estimate it by using the vev hierarchy $v_\rho \ll v_\chi \ll v_\sigma \simeq v_\varphi (\simeq\langle \phi_r\rangle$) assumed throughout this work. At leading order, the fields which get masses proportional to the largest scale $v_\sigma \simeq v_\varphi$ control $B$. In the scalar sector, the only field with mass proportional to $v_\varphi$ is $H_2$, and its mass, obtained from Eq. (\ref{MS}), can be written as $m_{H_2} \simeq \lambda_{H_2}^{1/2} v_\varphi$, where $\lambda_{H_2}$ is the effective coupling. In the vector boson sector, there is also only one super heavy field: $Z^{\prime \prime}$, whose mass is given in Eq. (\ref{zzmass}) and reads approximately $m_{Z^{\prime\prime}} \simeq 2 g_N v_\varphi$. Furthermore, many fields in the fermion sector need to be taken into account at leading order. As seen in Sec. \ref{subsecQuarks}, there is a total of 9 quarks fitting this criterion - 3 exotic quarks: $q_a^{(-4/5)}$ and $q^{(5/3)}$, as well as 2 up- and 4 down-type quarks. Similarly, for the neutral leptons, we have seen that 6 of them have (Majorana) masses proportional to $v_\varphi$. The masses of these heavy fermions follow the same pattern and can be conveniently written as $m_{f_i} \simeq h_{f_i} v_\varphi/\sqrt{2}$. For the sake of simplicity, in what follows, we assume that all Yukawas are equal: $h_{f_i}=h_f$. Thus, considering all these heavy fields, the leading contribution to $B$ becomes
\begin{eqnarray}\label{B2}
B\simeq \frac{1}{64\pi^{2}}\left(\lambda^2_{H_2}+48 g_N^4 - 30 h_f^4\right).
\end{eqnarray}
In this limit, the constraint $B>0$ can be easily satisfied if, for instance, $\lambda_{H_2}$, $g_N$ and $h_f$ are of the same order. If, for instance, such dimensionless parameters are of order 0.1, then all of the associated fields are super heavy -- the scalar and vector boson mass contributions are $m_{H_2} \simeq 316$ TeV and $m_{Z^{\prime\prime}} \simeq 200$ TeV, respectively, while each heavy fermion mass is $m_{f_i} \simeq 71$ TeV -- leading to the following scalon mass $m_S\simeq 12$ TeV. Nevertheless, it is worth pointing out that scenarios where only one of the bosonic fields, $H_2$ or $Z^{\prime\prime}$, is heavy enough to ensure $B>0$ can be viable as well. In such cases, the other boson could be much lighter by adjusting $g_N$ or $\lambda_{H_2}$ -- in fact, even lighter than the scalon field -- since its mass would be free from potential stability constraints.

\section{Conclusions}\label{sec:conc}

We have proposed a scale-invariant extension of the SM based on the $\text{SU}(3)_C\otimes \text{SU}(3)_L\otimes \text{U}(1)_X\otimes \text{U}(1)_N$ gauge group. Besides exhibiting the appealing features of the well-known 3-3-1 models, the construction includes an elegant realisation of a gauged $B-L$ symmetry. The latter symmetry arises naturally as a combination of one of the diagonal generators of SU$(3)_L$ and the generator of U$(1)_N$, in analogy with the electric charge. The $B-L$ symmetry is promoted to local and its breaking is triggered by the vev of a scalar singlet. An attractive consequence of the $B-L$ symmetry is that it leads to simplified textures for the fermion mass matrices, allowing for the direct implementation of seesaw mechanisms in this sector. In addition, these textures prevent potentially dangerous large mixing between the SM model and the new fermions.

Assuming a scalar sector composed of two $\text{SU}(3)_L$ triplets plus two singlets, a complex and a real, we have studied the one-loop effective potential and shown that symmetry breaking can be triggered dynamically. We have shown that all the conditions required by the Gildener-Weinberg method are satisfied. In particular, we have demonstrated that the tree-level stability conditions arising from the copositivity criteria are automatically satisfied when one requires that all the scalar masses and the Hessian matrix are positive semidefinite.

With most of the scalar degrees of freedom absorbed by the gauge sector, the scalar spectrum turned out to be minimal, featuring the electro-weak Higgs boson plus three other CP-even neutral fields, $H_1$, $H_2$ and the scalon $S$, with masses around or above the TeV scale, as well as a singly charged TeV-scale scalar, $H^\pm$. In the extended gauge sector, we have identified the SM gauge bosons -- $\gamma$, $Z$ and $W^{\pm}$ -- as well as new bosons, some of which become massive after the breaking of the $\text{SU}(3)_L\otimes \text{U}(1)_X$ symmetry by $v_\chi =\mathcal{O}(10)$ TeV -- extra (real) neutral and charged fields $W^{\pm\prime}$, $Z^\prime$ and a complex neutral boson $Y^{0(\dagger)}$ -- and a heavier neutral boson $Z^{\prime\prime}$ associated with the breaking of the $B-L$ symmetry, as defined in Eq. (\ref{BmL}), by $v_\sigma = \mathcal{O}(10^3)$ TeV.

Regarding the fermion sector, we have seen that, with a minimal field content, neutrinos and some generations of quarks remain massless. Nevertheless, we have presented how this drawback can be circumvented with the introduction of vectorial multiplets of quarks and neutral leptons, which become heavy due to their interactions with the scalar singlets, whose vevs ($v_\sigma\simeq v_\varphi$) are the largest energy scales in the model. As a consequence, all quarks become massive, with the previously massless fields getting their masses via seesaw mechanisms mediated by the new, heavy quarks. Thus, mass hierarchies arise among quark families, partly explaining the flavour puzzle. Similarly, neutrino masses are generated via a (double) seesaw mechanism involving the new (and heavy) neutral leptons. Finally, as our construction only explains partially the mass hierarchy observed in the quark sector -- leaving {\it e.g.} the down quark mass unsuppressed --  and does not explain the hierarchy among the charged lepton masses, we believe that such issues are worth exploring in future studies.

\acknowledgments
A. G. Dias thanks Conselho Nacional de Desenvolvimento Cient\'{\i}fico e Tecnol\'ogico (CNPq) for its financial support under the grant 305802/2019-4. J. Leite acknowledges financial support via grants 2017/23027-2 (FAPESP), PID2020-113775GB-I00 (AEI/10.13039/501100011033) and CIPROM/2021/054 (Generalitat Valenciana). B. L. S\'anchez-Vega would like to thank CNPq for its financial support under the grant 311699/2020-0.

\bibliographystyle{apsrev4-1}

\bibliography{3-3-1}

\end{document}